\documentclass[a4paper,11pt,notitlepage]{article}
\usepackage{etoolbox}
\usepackage{comment}
\makeatletter
\patchcmd{\chapter}{\if@openright\cleardoublepage\else\clearpage\fi}{}{}{}
\makeatother
\newcommand*\diff{\mathop{}\!\mathrm{d}}
\DeclareSymbolFont{matha}{OML}{txmi}{m}{it}
\DeclareMathSymbol{\varv}{\mathord}{matha}{118}

\def\app#1#2{%
  \mathrel{%
    \setbox0=\hbox{$#1\sim$}%
    \setbox2=\hbox{%
      \rlap{\hbox{$#1\propto$}}%
      \lower1.1\ht0\box0%
    }%
    \raise0.25\ht2\box2%
  }%
}

\usepackage{setspace}
\usepackage{caption}
\usepackage{subcaption}
\usepackage{natbib}
\usepackage{amssymb}
\usepackage{pifont}
\usepackage{bbm}
\usepackage{dsfont}

\usepackage{amsfonts}
\usepackage{amsmath}

\usepackage{listings}
\usepackage{authblk}
\usepackage{hyperref}
\usepackage{graphicx}
\usepackage{adjustbox}
\usepackage[toc,page]{appendix}
\usepackage{wrapfig}
\usepackage{titling}
\usepackage{float}
\usepackage[nodayofweek,level]{datetime}

\usepackage{xr}
\usepackage{mwe}
\usepackage[margin=1in]{geometry}
\usepackage{listings}
\usepackage[usenames,dvipsnames]{color}
\usepackage[UKenglish]{babel}
\usepackage[useregional]{datetime2}
\newcommand{\mydate}{\DTMdisplaydate{2019}{7}{15}{-1}}
\title{Ranking, and other Properties, of Elite Swimmers using Extreme
Value Theory}
\author{Harry Spearing$^1$, Jonathan Tawn$^1$, David Irons$^2$, Tim Paulden $^2$, Grace Bennett$^2$\\
{\small $^1$ Lancaster University, $^2$ ATASS sports} }

\date{\mydate}
\affil{Lancaster University}
\begin{document}
\maketitle

\begin{abstract}
The International Swimming Federation (FINA) uses a very simple points system with the aim to rank swimmers across all swimming events. The points acquired is a function of the ratio of the recorded time and the current world record for that event. With some world records considered ``better'' than others however, bias is introduced between events, with some being much harder to attain points where the world record is hard to beat. A model based on extreme value theory is introduced, where swim-times are modelled through their rate of occurrence, and with the distribution of the best times following a generalised Pareto
distribution. Within this framework, the strength of a particular swim is judged based on its position compared to the whole distribution of swim-times, rather than just the world record.
This model also accounts for the date of the swim, as training methods improve over the years,
as well as changes in technology, such as full body suits. The parameters of the generalised Pareto distribution, for each of the 34 individual long course events, will be shown to vary with covariates, leading to a novel single unified description of swim quality over all events and time. This structure, which allows
information to be shared across all strokes, distances, and genders, improves the predictive power as well as the model robustness compared to equivalent independent models. A by-product of the model is that it is possible to
estimate other features of interest, such as the ultimate possible time, the distribution of
new world records for any event, and to correct swim-times for the effect of full body suits. The methods will be illustrated using a dataset of the best 500 swim-times for each event in the period 2001-2018.
\end{abstract}


\section{Introduction}
On the face of it, comparing the performances of two swimmers in a given competition appears straightforward, simply compare their swim-times. But this simple comparison no longer holds when we compare between different distances, strokes or genders, let alone swimmers under different regulations for full body suits. In addition, due to the improvement in training methods, as well as changes in technology, a fair comparison between swim-times recorded many years apart is infeasible without some adjustment for the era of the swim. 

The International Swimming Federation (FINA) uses a very simple points system to tackle this issue. The points acquired for a particular swim is a function of the ratio of the swim-time and the current world record for that event, specifically the points $p_{i,j}$ given to swimmer $i$ in event $j$ is $p_{i,j} \propto (b_j/t_{i,j})^3$ where $b_j$ is the current world record in event $j$, and $t_{i,j}$ is the time of swimmer $i$ in event $j$. With some world records considered better than others however, bias is introduced between events, with some being much harder to attain points where the world record is hard to beat. Furthermore, the ranking method has high sensitivity as it is determined only by the set of current world records, so rankings can change substantially when a single record is broken. Importantly, FINA rankings are used by many countries and organisations for selection for regional and international competitions, so the ranking must be an accurate representation of the swimmer's true ability. 

The aim, is to produce a global model that can fairly compare between strokes, gender and distance, as well as considering the improvement over time of elite sporting performance. This paper utilises extreme value theory to model the very best swim-times as being observations from a generalised Pareto distribution (GPd) so that the strength of a particular swim is judged on its position compared to the whole distribution of swim-times across all events, rather than just the world record for that event. This ensures a more efficient comparison between events. Moreover, comparison between swim-times within the same event has a more tangible interpretation since it can be described in terms of probabilities. For example, by considering swim-times $t_1$ and $t_2$ in an event, then it is natural to compare the relative quality of these by $\Pr(T>t_1)/\Pr(T>t_2)$, where $T$ represents the random variable corresponding to a swim-time for an event, rather than, say, the metric $t_1-t_2$.
A by product of this global model is that other features of interest can be estimated, for example the ultimate possible swim-time for any given event. The distribution of the next world record swim-time for each event can be estimated, and even the distribution of the waiting time, and therefore the expected waiting time, until the next world record is broken and the probability of that record being in a particular event. In addition, swim-times can be corrected for the effect of full body suits, to allow for fair comparison between those swimmers wearing suits and those not.

The data to be studied comprise the top 500 swim-times, with at most one time per swimmer per event, in all 34 individual long course (LC) swimming events, i.e., in a 50m pool, from all major competitions between the start of 2001 and last quarter of 2018. Any data not officially accepted by FINA are removed, for example observations that were later rescinded due to the use of performance enhancing drugs. For the remainder of this article, \textit{negative swim-times} will be analysed, and simply referred to as \textit{swim-times}, so that if a swim-time is faster than another it has the larger negative swim-time of the two. So, for the best swim-times we are interested in the biggest negative swim-times. Therefore the paper focuses on methods for largest values, which is the typical methodological approach to extreme values \citep{coles2001introduction}. Results for actual swim-times are obtained by simply negating the results we obtain for negated swim-times. Additionally, independence is assumed between all swim-times across different years, strokes and distances, even if they are achieved by the same swimmer. Both of these two points will be discussed further in Section \ref{sec:Discussion}.

The past use of extreme value theory for sports modelling is varied. In athletics, work has been done to create a model which pools information between different distances and over time \citep{stephenson2013determining}. The threshold exceedance model of \cite{smith1989extreme} is used by \cite{strand1998modeling} to model times of long distance runners. Specifically, the typical change of time taken to run 10 kilometres with respect to the age of the athlete is modelled via a Gumbel distribution, where times within ages and across ages are assumed to be independent, and men's and women's times are modelled separately. More generally, \cite{riegel1981athletic} finds a linear relationship between log world record time and log distance over many sports. Modelling men's and women's data separately is a common theme in sports data.

In swimming, \cite{gomes2019swimming} use extreme value theory to model the distribution of swim-times across all LC events using independent fits for each event. \cite{doi:10.1080/15598608.2012.695702} explore the progression of the top performances in swimming events over time by modelling the times of the gold medallist swimmers in the Olympic Games. Dependence due to the same swimmer winning two events at an Olympic Games is included via a bivariate extreme value distribution \citep{tawn1988bivariate}. We are unaware of any previous publication that models swim data globally across gender, distance, stroke, and considers the improvements over time.	

The article is set out as follows. Section \ref{sec:Theory} introduces extreme value theory, and the point process representation of \cite{smith1989extreme}, see also \cite{coles2001introduction}, which forms the basis of our model. Section~\ref{sec:Modelling} describes the full global model and the justification for the shared fit. In Section~\ref{sec:results} the features of interest discussed above will be estimated based on the final fitted model, such as the ultimate possible swim-time for each event, examples of the best swimmers of all time under this model, the distribution of new world records, the expected time until the next world record is broken, the probability of the next world record being in a given event, and the result of adjusting for regulations of full body suits on current world records. Section~\ref{sec:Discussion} discusses the possible impacts of any major assumptions made in the modelling process, as well as investigating further improvements and applications to the proposed model.

 \section{Theory}
 \label{sec:Theory}

 \subsection{Extremes of identically distributed variables}

\begin{figure}
\centering
\begin{minipage}{.45\textwidth}
  \centering
  \includegraphics[scale=0.45]{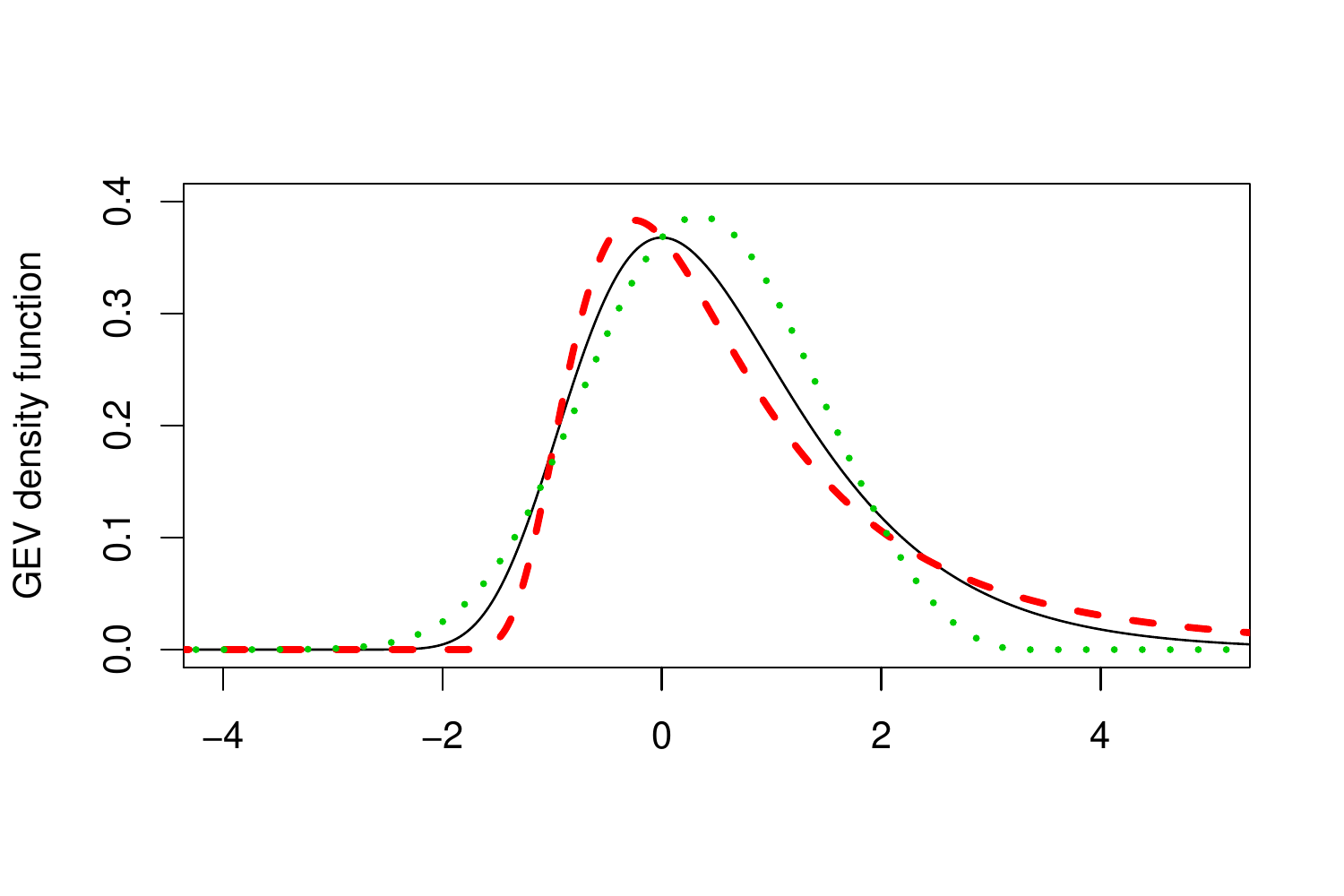}
  
\end{minipage}%
\begin{minipage}{.05\textwidth}
\hspace{1mm}
\end{minipage}
\begin{minipage}{.45\textwidth}
  \centering
  \includegraphics[scale=0.45]{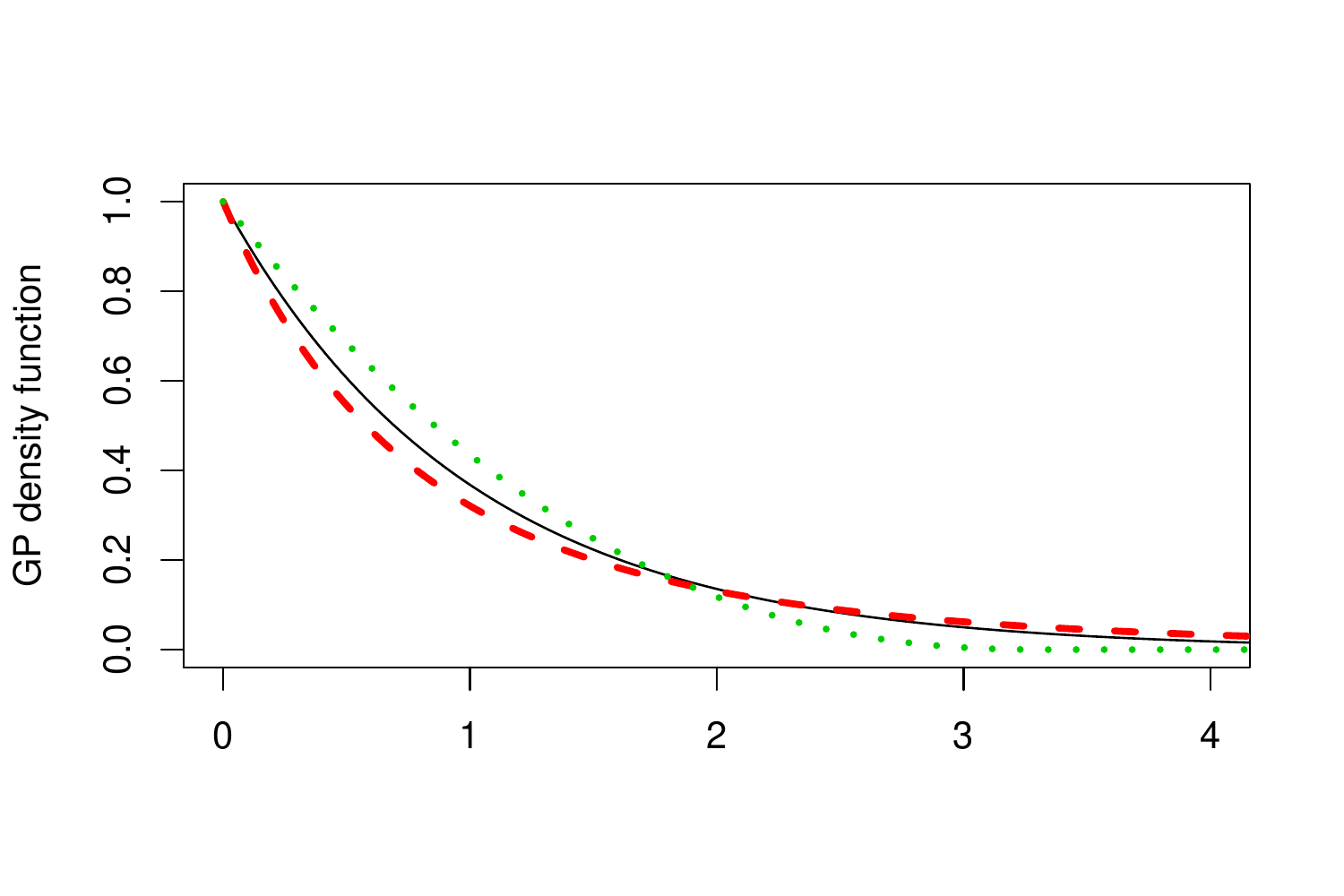}

\end{minipage}
\caption{Density functions for the 
	GEVd $(\mu=0, \sigma=1, \xi)$ (left) and GPd $(u=0, \tilde{\sigma}_u=1, \xi)$ (right) for three different shape parameters: $\xi = 0$ (solid line), $\xi = -0.3$ (dotted line) and $\xi=0.3$ (dashed line).}
	\label{fig:densityGEVd}
\end{figure}

Univariate extreme value theory (EVT) provides the framework for our modelling strategy. In its simplest form it applies to an independent identically distributed (IID) random sample $X_1,\dots, X_n$ with each variable having a continuous distribution function $F$. The two main approaches in EVT are the block maxima method and the peaks over threshold methods. The asymptotic theory behind these two methods is as follows. Let $M_n= \max\{X_1,\dots, X_n\}$ be the maximum of a block of length $n$. We seek the distribution of $M_n$ for large $n$, and in particular appropriate choices of norming sequences $a_n > 0$ and $b_n$ are sought such that, as $n\rightarrow\infty$,
\begin{eqnarray}
\Pr\left\{\frac{M_n-b_n}{a_n} \leq x \right\} & = & \Pr(X_1 \leq a_nx +
 b_n, \dots, X_n \leq a_nx+b_n) \nonumber \\
 & = & F^n(a_n x + b_n) \nonumber \\
 & \rightarrow & G(x)
\label{eq:probMax}
\end{eqnarray}
where the limiting distribution $G(x)$ is non-degenerate. The only possible non-degenerate limiting distribution of equation \eqref{eq:probMax} is the generalised extreme value distribution function (GEVd). The exact form is given by
\begin{equation}
G(x) = \exp\left(-[1+\xi(x-\mu)/\sigma]^{-1/\xi}_+\right),
\label{eq:GEVd}
\end{equation} 
 where $\mu,\; \xi \in \mathbb{R}, \;\sigma \in \mathbb{R}^+$, are the location, shape and scale parameters respectively and $ y_+ = \max(y,0).$ 
Figure~\ref{fig:densityGEVd} (left) illustrates the density of the GEVd for different values of $\xi$, while $\mu=0$, $\sigma=1$. For $\xi < 0$, there exists a finite value $x_G = \mu - \sigma/\xi: \; G(x) = 1, \; \forall x>x_G $. In contrast, for $\xi\geq 0, \; G(x) < 1, \; \forall x<\infty$. The GEVd result is powerful as it holds as the limit distribution for a very broad class of continuous distributions $F$ and implies that whatever $F$ is in this class, the maxima must follow a single class of distributions, determined by only three parameters.

The block maxima method of \cite{coles2001introduction} assumes that limit~\eqref{eq:probMax} holds exactly for a large enough block size $n$, for example all observations in a month or a year. Given a sample of length $kn$ the approach is to split the series into $k$ blocks with $n$ values in each block. Then the $k$ values of the block maxima are used to estimate the parameters $(\mu, \sigma, \xi)$ of the model, assuming that each of these variables is IID and follows a GEVd.

The peaks over threshold (POT) approach considers only the observations above a suitably high threshold. This allows all of the most extreme data to be analysed, unlike the block maxima approach, and typically leads to more efficient inference.
Let $$N_n(x) = \sum_{i=1}^n \mathds{1}\left( X_i > a_n x + b_n \right), $$ with $\mathds{1}(A)$ be an indicator of event $A$ occurring, then $N_n(x)$ is the random variable corresponding to the number of $X_1,\dots,X_n$ exceeding a threshold $a_n x + b_n$, with $a_n$ and $b_n$ as in limit~\eqref{eq:probMax}. So $N_n(x)$ has a Binomial distribution with $N_n(x) \sim \operatorname{B}(n, 1 - F(a_n x + b_n))$. Under the same conditions behind the GEVd limit 
from equation \eqref{eq:probMax}, as $n \rightarrow\infty$,
\begin{equation*}
n \log F(a_n x + b_n) \rightarrow \log G(x),
\end{equation*}
and so, using standard Taylor series approximation, for all $x$
 \begin{equation}
 \label{eq:gpd2}
 n[1-F(a_n x + b_n)] \rightarrow -\log G(x) = [1+\xi(x-\mu)/\sigma]^{-1/\xi}_+.
 \end{equation}
Using property \eqref{eq:gpd2}, then the classic Poisson limit from a Binomial gives that as $n~\rightarrow~\infty$, $N_n( x )~\rightarrow~N(x)$, where $N(x)$ is a Poisson random variable with mean $[1+~\xi(x-~\mu)/\sigma]^{-1/~\xi}_+$. Furthermore, it follows that for $x>u$ and $X$ distributed as $X_i$, that

\begin{equation}
\Pr\{X > a_n x + b_n | X > a_n u + b_n\} \rightarrow \log G(x)/ \log G(u) = \bar{H}_u(x),
\label{eq:probExc}
\end{equation}
where $\bar{H}_u(x) = 1-H_u(x)$, where the distribution function $H_u$ is of the form
\begin{equation}
H_u(x) \equiv 1-\left[1+\xi \left(\frac{x-u}{\tilde{\sigma}_u}\right)\right]^{-\frac{1}{\xi}}_+,
\label{eq:GPd}
\end{equation}
is the generalised Pareto distribution function (GPd) with threshold $u$, shape parameter $\xi$ and 
scale parameter $\tilde{\sigma}_u \in \mathbb{R}^+$ is linked to the GEVd parameters via $\tilde{\sigma}_u = \sigma + \xi(u-\mu)$. Limit distribution $H_u$ gives an asymptotic model for the distribution of exceedances above a threshold $u$, no matter what the distribution $F$. Figure \ref{fig:densityGEVd} (right) illustrates the density of the GPd for different values of $\xi$. For $\xi < 0$, there exists a finite value $x_H = u - \tilde{\sigma}_u/\xi: \; H_u(x) = 1, \; \forall x>x_H $. In contrast, for $\xi\geq 0, \; H_u(x) < 1, \; \forall x<\infty$.

The POT approach leads to a model for the extreme tail with two components: a model for the number of exceedances of the threshold, which is Poisson with mean $\lambda = [1+\xi(x-\mu)/\sigma]^{-1/\xi}_+$, and a model for threshold exceedances, $H_u(x)$ which is a GPd. The choice of threshold $u$ is user-specified, with the choice based on the usual bias-variance trade-off, the subject of much historical focus \citep{scarrott2012review}. 

As can be seen from the derivation above both the rate and GPd parameters are functions of the GEVd parameters. In fact, the block maxima and POT approaches can be combined using a point process limit which exploits this property.
Consider, the point process model of extremes, defined on a sequence 
\begin{equation*}
P_n = \left\{\left(\frac{i}{n+1}, \frac{X_i-b_n}{a_n}\right) \; : \; i=1,\dots, n\right\},
\end{equation*}
where the scaling here enforces that, as $n\rightarrow \infty$, the first component is continuous on $[0,1]$, and the maximum of the second component to be non-degenerate with limiting distribution \eqref{eq:GEVd}. In particular as $n \rightarrow \infty$, $P_n \rightarrow P$ where $P$ is a non-homogeneous Poisson process on
$(0,1] \times (b_l, \infty)$, where $b_l=\max\{x\in \mathbb{R}: G(x)=0\}$ where $G$ is the limit distribution~\eqref{eq:GEVd} \citep{smith1989extreme}.
It follows that the integrated intensity $\Lambda$ of $P$ on $\mathcal{A}_{t,x} =[0,t] \times [x, \infty]$, where $0 < t \leq 1 , \; x>b_l$ is
\begin{equation*}
\Lambda \left(\mathcal{A}_{t,x}\right) = t \left[1+\xi \left(\frac{x-\mu}{\sigma}\right)\right]^{-\frac{1}{\xi}}_+,
\end{equation*}
which implies that the intensity function $\lambda$ for $P$ is, for $t\in (0,1]$ and $x>b_l$, 
\begin{equation}
\label{eq:intensityDens}
\lambda(t,x) =  \frac{\partial ^2  \Lambda\left(\mathcal{A}_{t,x}\right)}{\partial x \; \partial t}  =  \frac{1}{\sigma}\left[1 + \xi \left(\frac{x - \mu }{\sigma}\right)\right]_{+}^{-\frac{1}{\xi}-1}.
\end{equation}
From standard Poisson process properties we have that the number of points of $P$ in any set $S\subseteq [0,1] \times (b_l, \infty)$ follows a Poisson distribution with mean $\Lambda(S)=\int_S \lambda(t,x) \diff x \diff t$, with $\lambda(t,x)$ given by expression~\eqref{eq:intensityDens}.

Statistical application of the point process model assumes that for large enough $n$, the limit $P_n\rightarrow P$ holds exactly. After absorbing norming constants into the limiting intensity, it is assumed that $P$, with intensity~\eqref{eq:intensityDens}, applies to the points $\{(i/(n+1),X_i); i=1, \ldots ,n\}$ on the set $\mathcal{A}_{1,u}=[0,1] \times (u, \infty]$.
If $\pmb{x}=\{(t_1,x_1),\dots,(t_m,x_m)\}$ denote the $m$ of these points that fall in $\mathcal{A}_{1,u}$, then
the likelihood for the parameters $\theta = (\mu, \sigma, \xi)$ is  
\begin{equation}
\label{eq:point_process}
L(\theta ; \pmb{x}) = \exp\left\{-\Lambda(\mathcal{A}_{1,u})\right\} \prod_{i=1}^m \lambda(t_i,x_i).
\end{equation} 
Inference using this likelihood gives information about both the mean number of exceedances of the threshold $u$ and the distribution of the threshold exceedances (the GPd). When a datum $x_i$ has been recorded to some precision $s$ such that the true value $x'_i$ is unknown but $x'_i \in [x_i-s/2, x_i+s/2)$,  interval censoring is introduced, which can be factored into the likelihood via 
\begin{align*}
L(\theta ; \pmb{x})& \propto \exp\left\{-\Lambda(\mathcal{A}_{1,u})\right\} \prod_{i=1}^m \int_{x_i-s/2}^{x_i+s/2}\lambda(t_i,x) \diff x \nonumber\\
& =  \exp\left\{-\Lambda(\mathcal{A}_{1,u})\right\} \prod_{i=1}^m \left\{\left[1+ \xi \left(\frac{x_i - s/2 - \mu}{\sigma} \right)\right]^{-\frac{1}{\xi}}_+ - \left[1+ \xi \left(\frac{x_i + s/2 - \mu}{\sigma}\right) \right]^{-\frac{1}{\xi}}_+ \right\}.
\end{align*}

\subsection{Extreme values of non-identically distributed variables}
\label{sec:nonidentically}
The derivations so far have assumed IID variables, however this need not be the case. Whilst still assuming independence, the assumption of identically distributed data is relaxed by including a covariate structure. In order to take the date of the swim into consideration, time is introduced as a covariate such that,
in the most general case, all parameters of $\theta$ are allowed to vary with time, for example $\theta(t) = (\mu(t),\sigma(t),\xi(t))$. The non-homogeneous Poisson process allows for time-dependent rates of occurrences and excess distributions, see \cite{smith1989extreme}. Under this relaxation, equation~\eqref{eq:intensityDens} becomes 
\begin{equation}
\lambda(t,x) =  \frac{1}{\sigma(t)}\left[1 + \xi(t) \left(\frac{x - \mu(t) }{\sigma(t)}\right)\right]^{-\frac{1}{\xi(t)}-1}_+,
\label{eq:intensityDensT}
\end{equation}
and so the integrated intensity is
\begin{equation}
\Lambda(\mathcal{A}_{1,u}) = \int_0^1 \left[1+\xi(t) \left(\frac{u-\mu(t)}{\sigma(t)}\right)\right]^{-\frac{1}{\xi(t)}}_+ \diff t.
\label{eq:IntensityT}
\end{equation}
The full likelihood function, accounting for interval censoring, can then be expressed, as in equation \eqref{eq:point_process}, but with $\Lambda\left(\mathcal{A}_{1,u}\right)$ and $\lambda(t,x)$ given in equations \eqref{eq:IntensityT} and \eqref{eq:intensityDensT}, such that
\begin{align}
\label{eq:likelihood_t}
L(\theta ; \pmb{x}) &= \exp\left\{-\Lambda(\mathcal{A}_{1,u})\right\} \prod_{i=1}^m \int_{x_i-s/2}^{x_i+s/2}\lambda(t_i,x) \diff x \nonumber \\
&=  \exp\left\{-\Lambda(\mathcal{A}_{1,u})\right\} \prod_{i=1}^m \left\{\left[1+ \xi(t_i) \left(\frac{x_i - s/2 - \mu(t_i)}{\sigma(t_i)} \right)\right]^{-\frac{1}{\xi(t_i)}}_+ - \right.  \nonumber \\
 & \qquad \qquad \qquad \qquad \qquad \qquad \left. \left[1+ \xi(t_i)  \left(\frac{x_i + s/2 - \mu(t_i)}{\sigma(t_i)}\right) \right]^{-\frac{1}{\xi(t_i)}}_+ \right\},
\end{align}
 where the parameters within $\theta(t)$ are found by maximising this likelihood. If $\{y_i: i=1,\dots,18\}$ is the set of start dates of years from 2001-2019, then the expected rate of exceedances of $u$ with year $2000+i$ is given by 
\begin{equation*}
\Lambda_i(\mathcal{A}_{1,u}) = \int_{y_i}^{y_{i+1}} \left[1+\xi(t) \left(\frac{u-\mu(t)}{\sigma(t)}\right)\right]^{-\frac{1}{\xi(t)}}_+ \diff t.
\end{equation*}
If the change in the parameters is small over the course of each year, then the rate can be approximated as \begin{equation}
\label{eq:rate_approx}
\Lambda_i(\mathcal{A}_{1,u}) \approx 
\left[1+\xi(y^*_i) \left(\frac{u-\mu(y^*_i)}{\sigma(y^*_i)}\right)\right]^{-\frac{1}{\xi(y^*_i)}}_+ (y_{i+1} - y_i),
\end{equation} where $y^*_i = (y_i + y_{i+1})/2$. Likewise the excess distribution at a time $t$ is given for $x> u$ by
\begin{equation*}
\Pr\{X_t > x | X_t > u\} =  \left[1 + \xi(t) \left(\frac{x - u }{\tilde{\sigma}_u(t)}\right)\right]^{-\frac{1}{\xi(t)}}_+,
\end{equation*}
where $\tilde{\sigma}_u(t) = \sigma(t) + \xi(t)\left[u-\mu(t)\right]$.



\section{Model for swimming data}
\label{sec:Modelling}
\subsection{The Data}
\label{sec:thedata}
The data are from the FINA swimming website's database, at \url{http://www.fina.org/}, which contains around the top 500 recorded swim-times for all 34 individual LC swimming events. The fastest swim time per swimmer per event is taken, irrespective of the year in which it occurs.
The data includes interval censored observations which come from the rounding of recorded timings. Given that the data are rounded, in seconds to 2 decimal places, the interval censoring likelihood~\eqref{eq:likelihood_t} is formally needed with $s=0.01$. In practice using standard likelihood \eqref{eq:point_process} instead would give similar results in practice, with the exception of 50m events as the rounding is a more substantial part of the variation in these data.

In order to develop a consistent approach across all events $e \in E$ where $E$ is the set of all 34 individual LC swim events, the threshold for each event was set such that there were an identical number of exceedances in each event. From plotting PP and QQ plots for each event $e$ independently over a range of thresholds $u'_e$, the thresholds were set such that there were 200 exceedances in each event, as this appropriately balances the bias and variance for the majority of events. For each event $e$, the threshold used for the model, $u_e$, was set to $u_e = u_e' -s/2$, to account for the interval censoring.

\begin{figure}
\centering
\begin{minipage}{.45\textwidth}
  \centering
  \includegraphics[scale=.5]{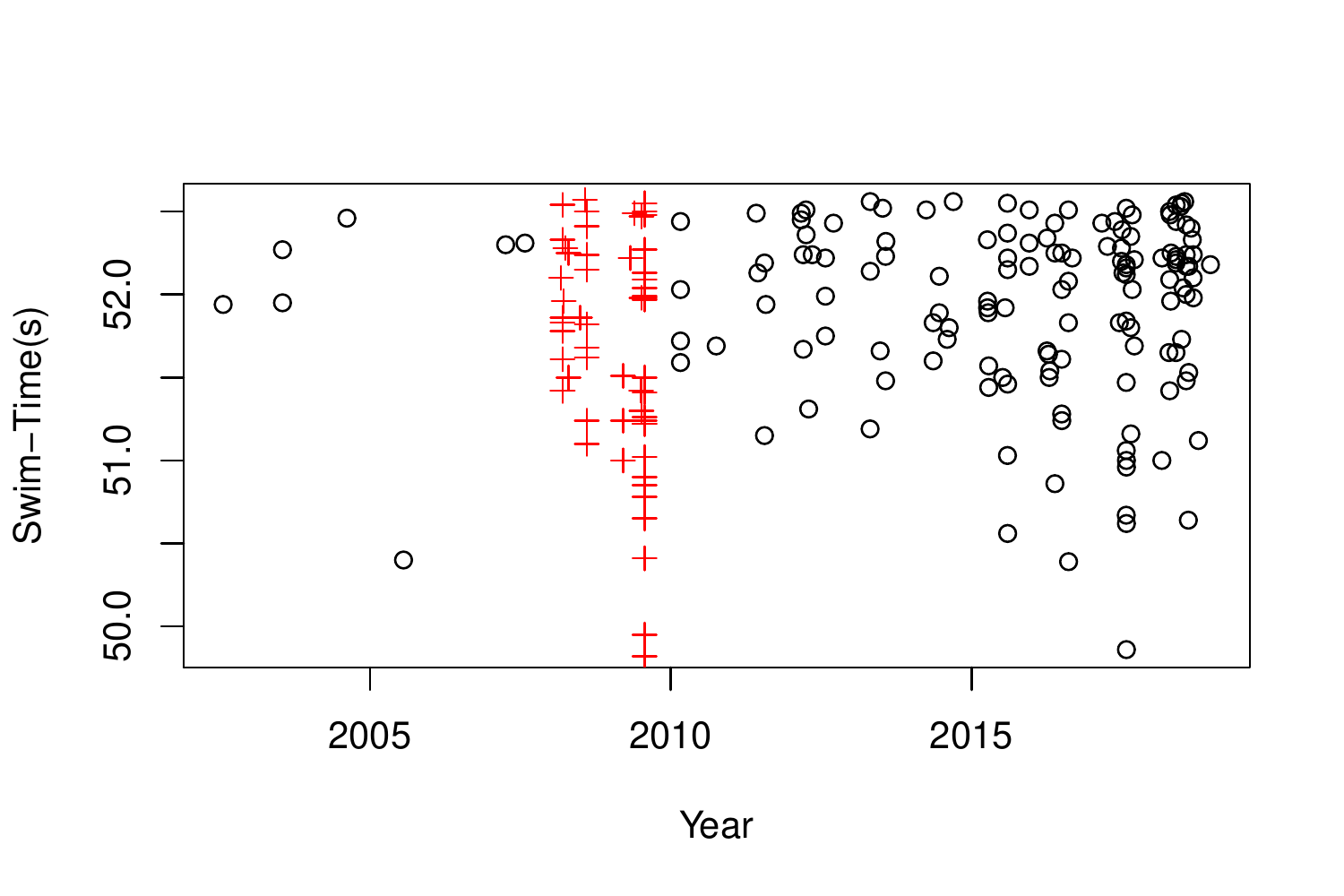}

\end{minipage}%
\begin{minipage}{.05\textwidth}
\hspace{1mm}
\end{minipage}
\begin{minipage}{.45\textwidth}
  \centering
  \includegraphics[scale=.5]{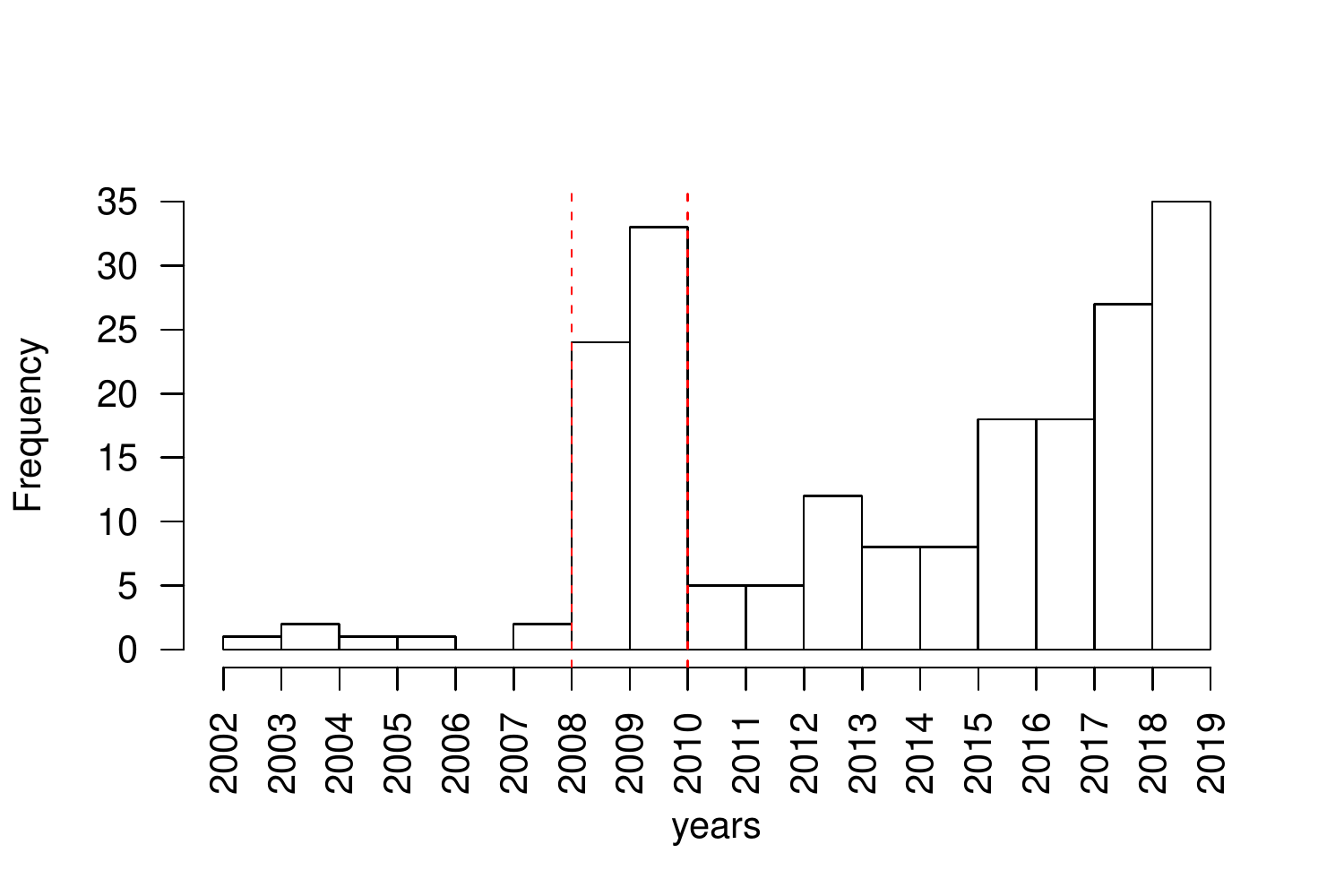}

\end{minipage}
\caption{Data for the men's 100m butterfly. The data (left) shows the raw data for the swim-times and so the lower tail is the feature of interest. Here, the crosses indicate swims recorded within then swim-suit period.  Similarly, the observed annual rates of exceeding the threshold (right) include dashed vertical lines (right) which indicate the swim-suit time period.}
\label{fig:Hist6}
\end{figure}

Properties of the 200 best times for the 100m men's butterfly swim-times are illustrated in Figure \ref{fig:Hist6}, with these being typical across all events. There is a
 general increasing trend in the rate of occurrences over time. In addition to this trend there is a noticeable step-increase in the frequency of observations in the top 200 swims between the introduction, in 2008, and subsequent banning, from the start of 2010, of \textit{swim-suits} by FINA \citep{shipley2009fina}. Swim-suits have been found to reduce drag by up to 35\% in independent testing \citep{moria2011aero}, and a significant number of world records were set during their use. Particularly in 2009, the introduction all polyurethane suits, such as the `Arena X-Glide', saw a significant improvement in performances \citep{foster2012influence}. Figure \ref{fig:Hist6} shows that there appears to be differences in performances between 2008 and 2009 which illustrates an impact of changes of full-body suit technology.

There is an inconsistency in the selection of the competitions in the FINA database, with only important competitions being represented in some of the earlier years, whereas later years cover all high-level competitions. One consequence of this is that the rate per year of exceeding the threshold $u_e$ will increase over time due to this feature, with the effect being largest in the earliest years. So, changes in the threshold exceedance rate, for each event, arise from a combination of improved swimming performance and the database formulation. Therefore, care must be taken when interpreting this feature in the analyses. There is also the potential for the distribution of swim times that exceed the threshold to change over time due to this biased selection of competitions in the database. Any such effect should be minimal on inferences given that most exceedances are from the later years, so the likelihood is naturally most influenced by data from later years. The model we develop presumes there is no such bias to the distribution of excesses, but this assumption is tested (see Figure \ref{fig:PPplot} (right)) and shown to provide a sufficiently good description of the early data.


\subsection{Separate Event Model}
The Poisson point process framework allows us to model the time varying rate of observations above threshold, as well as the distribution of these observations.
To incorporate the general increasing frequency of swim-times observed in Figure \ref{fig:Hist6}, time was included as a covariate in the model. The swim-suit factor was included via an indicator covariate, where the assumption is made that all observations during the swim-suit epoch were by swimmers wearing a swim-suit, and initially it is assumed that the swim-suit effect is constant throughout this epoch. 

Following \cite{davison1990models} and \cite{coles2001introduction} the Poisson process parameters $\mu^{(e)}(t)$, $\sigma^{(e)}(t)$ and $\xi^{(e)}(t)$ are initially assumed to vary smoothly with time $t$ in the model for each separate event. From fitting each event independently, it was then concluded, via use of AIC, that a linear dependence on time is appropriate for the parameters $\mu^{(e)}(t)$ and $\sigma^{(e)}(t)$ to describe the increase in rates of observations. Moreover, $\xi^{(e)}(t)$ is assumed to be constant over time as is common in the literature across extreme value applications to rainfall, sea-level, and athletics amongst others, e.g, \cite{smith1989extreme}, \cite{ robinson1995statistics}, \cite{strand1998modeling}, which find that despite changes in the distribution due to various covariates, the shape parameter is constant and is therefore taken as some unknown fixed value $\xi^{(e)}(t) = \xi^{(e)}$ for that event.

Although the patterns in the rate of observations is noticeable from plots alone, patterns in the distribution of the observations exceeding the threshold are not so obvious. To find an appropriate model for the distribution of exceedances above the thresholds, several models for the GPd parameters were fitted and compared, which included, but were not limited to, linear trends over time and including indicators of swim-suit effects. Interestingly, after model comparison it was concluded that for each event the distribution of observations above the threshold is independent of covariates, indicating that any improvements over time are due to an increase in quantity of exceedances above the threshold, rather than any change in the nature of the exceedances themselves. These findings in the data about the rate and the distribution of the best swims are reflected in the following parametrisations.

For a given event $e \in E$, the Poisson process is parametrised as either,
\begin{eqnarray}
&\xi^{(e)} (t)& = \xi^{(e)}
, \nonumber  \\
&\mu^{(e)}(t)& = \mu_0^{(e)} + \beta^{(e)} t + \gamma^{(e)} \mathds{1}_{\{t\in S_t\}}, \nonumber\\
&\sigma^{(e)}(t)& = \sigma_0^{(e)} + \xi^{(e)} \beta^{(e)} t + \xi^{(e)} \gamma^{(e)} \mathds{1}_{\{t\in S_t\}},
\label{eq:parameterisation}
\end{eqnarray}
or,
\begin{eqnarray}
&\xi^{(e)} (t)& = \xi^{(e)}
, \nonumber  \\
&\mu^{(e)}(t)& = \mu_0^{(e)} + \beta^{(e)} t + \gamma^{(e)}_1 \mathds{1}_{\{t\in S_{t_1}\}} + \gamma^{(e)}_2 \mathds{1}_{\{t\in S_{t_2}\}}, \nonumber\\
&\sigma^{(e)}(t)& = \sigma_0^{(e)} + \xi^{(e)} \beta^{(e)} t + \xi^{(e)} \gamma^{(e)}_1 \mathds{1}_{\{t\in S_{t_1}\}} + \xi^{(e)} \gamma^{(e)}_2 \mathds{1}_{\{t\in S_{t_2}\}},
\label{eq:parameterisation_twosuit}
\end{eqnarray}
where $\theta^{(e)}(t)$ represents $\theta$ for event $e$ at time $t$, and $\mu_0^{(e)}, \;\xi^{(e)} \in \mathbb{R}, \; \sigma_0^{(e)} \in  \mathbb{R}^+$ are the location, shape, and scale parameters for the Poisson process, $\beta^{(e)} \in \mathbb{R}$ controls the linear trend in $\mu^{(e)}(t)$ and $\sigma^{(e)}(t)$. In the case of assuming a single swim-suit effect, $\gamma^{(e)} \in \mathbb{R}$ controls the magnitude of this effect, $\mathds{1}$ is the indicator function and $S_t \in [2008, 2009]$ denotes the time period in which swim-suit were allowed, and in the case of allowing for the differing effects of the two major suit-types, as noted in Section \ref{sec:thedata}, $\gamma_1^{(e)} \in \mathbb{R}$ and $\gamma_2^{(e)} \in \mathbb{R}$ control the effects of these two suit-types, with $S_{t_1} \in [2008]$ and $S_{t_2} \in [2009]$ denoting the approximate time periods in which these suits were active. In particular $t$ is linearly standardised to have zero mean and unit variance over the observed data. Both parametrisation
s \eqref{eq:parameterisation} and \eqref{eq:parameterisation_twosuit} ensure that the GPd scale parameter for exceedances of the level $u_e$ at time $t$ is covariate-independent. For example, with parametrisation \eqref{eq:parameterisation},
\begin{align}
\label{eq:stationarysigma}
\tilde{\sigma}_u^{(e)} (t) &= \sigma^{(e)}(t) + \xi^{(e)}\left[u_e- \mu^{(e)}(t)\right]  \nonumber\\
&= \sigma_0^{(e)} + \xi^{(e)} \beta^{(e)} t \; + \; \xi^{(e)} \gamma^{(e)} \mathds{1}_{\{t\in S_t\}} + \xi^{(e)}(u_e - [ \mu_0^{(e)} + \beta^{(e)} t + \gamma^{(e)} \mathds{1}_{\{t\in S_t\}} ])  \nonumber\\
&= \sigma_0^{(e)} + \xi^{(e)} (u_e - \mu_0^{(e)}) \nonumber\\
&:= \tilde{\sigma}_u^{(e)},
\end{align} and the same clearly holds for parametrisation \eqref{eq:parameterisation_twosuit} so that the two GPd parameters, $\xi^{(e)}$ and $\tilde{\sigma}^{(e)}_u$, and thus the distribution above the threshold is identically distributed over covariates, as required. It is common to use a log link in the scale parameter in the non-homogeneous Poisson process to ensure positivity, however this would make the covariate independence of $\tilde{\sigma}_u^{(e)}$, property \eqref{eq:stationarysigma}, impossible. Instead, $\mu_0^{(e)}$, $\sigma_0^{(e)}$ and $\xi^{(e)}$ are constrained such that $\tilde{\sigma}^{(e)}_u$ in expression \eqref{eq:stationarysigma} is positive.

\begin{figure}
\centering
\begin{minipage}{.45\textwidth}
  \centering
  \includegraphics[scale=.5]{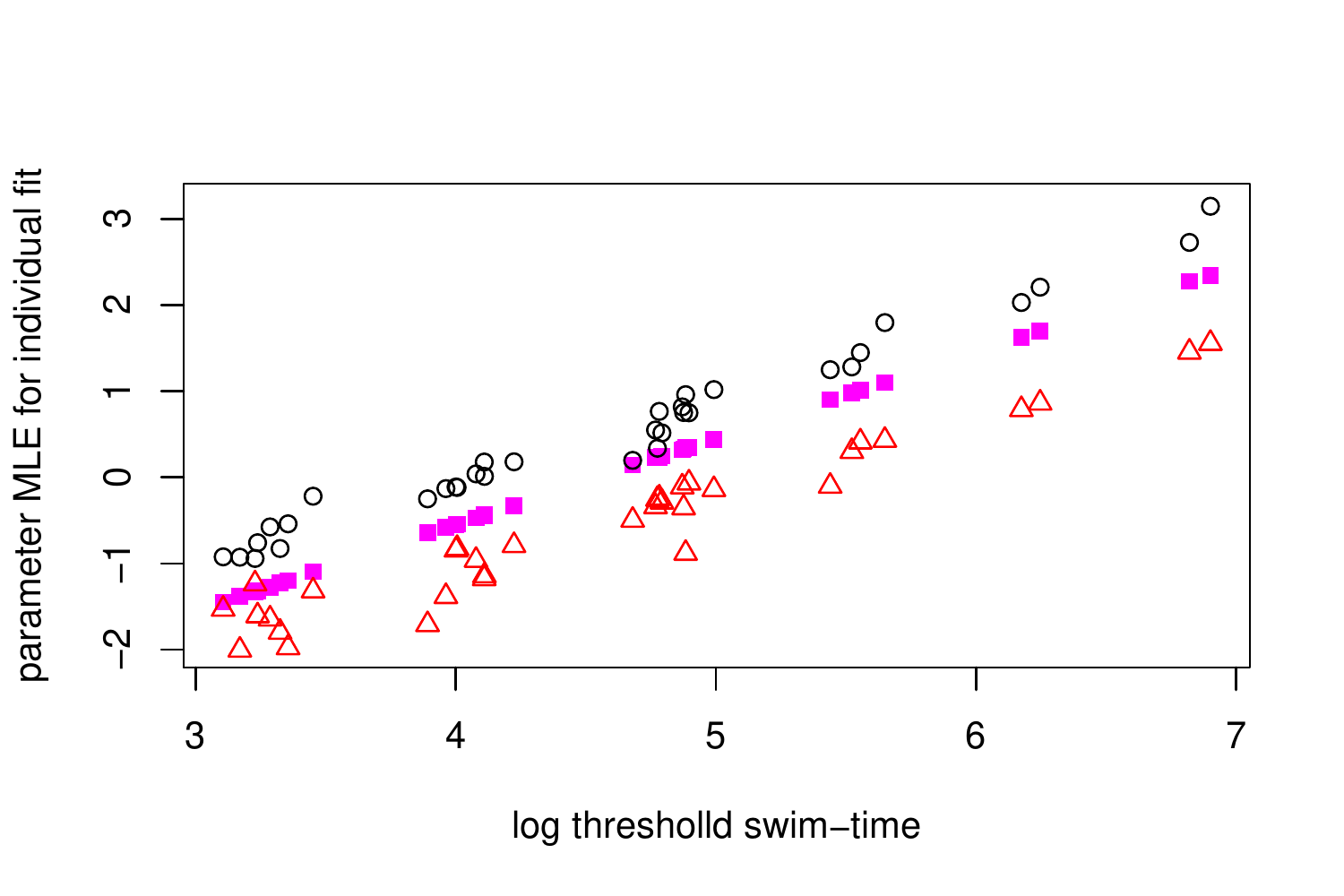}

\end{minipage}%
\begin{minipage}{.05\textwidth}
\hspace{1mm}
\end{minipage}
\begin{minipage}{.45\textwidth}
  \centering
  \includegraphics[scale=.5]{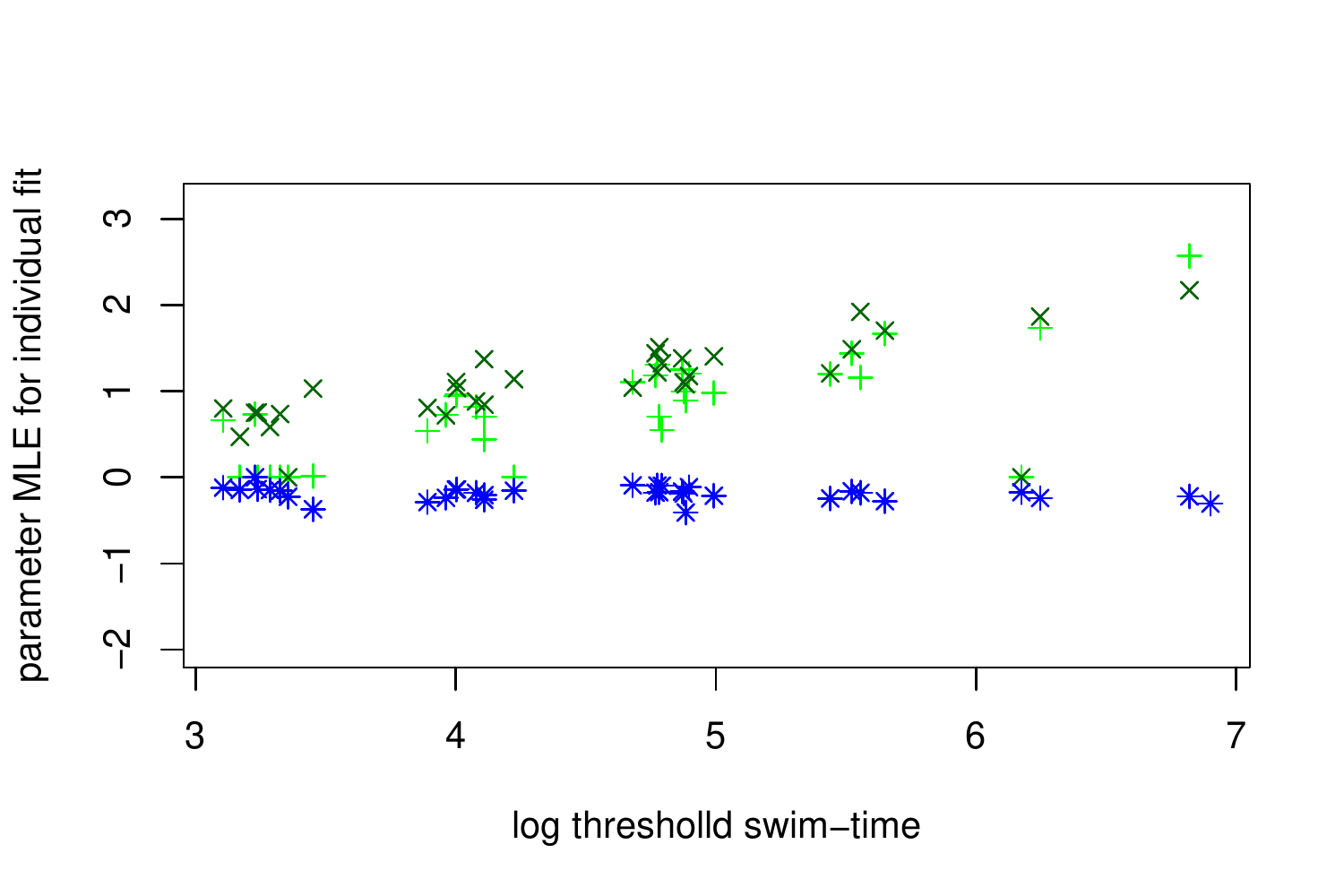}

\end{minipage}
\caption{Transformed parameter estimates against log threshold swim-time $u_L = \log(-u)$. A linear or near-linear relationship is apparent for most of the parameters: for $\sigma_L = \log(\tilde{\sigma}_u)$ (black circle), $\mu_L = \log(-\mu_0)$ (purple square), $\beta_L = \log(\beta)$ (red triangle),$\gamma_{L,1} = \sqrt{\gamma_1}$ (light-green plus, $+$) and $\gamma_{L,2} = \sqrt{\gamma_2}$ (dark-green cross, $\times$). The shape parameter $\xi$ (blue star) is approximately constant. Note that $\mu_L$ has been rescaled, by subtracting $5$ uniformly, to be visible on the plot.}
\label{fig:paramplots}
\end{figure}

Figure \ref{fig:paramplots} shows all the model parameter estimates from parametrisation \eqref{eq:parameterisation_twosuit}, obtained by fitting independently across events: three GEV parameters, $\mu_0$, $\sigma_0$, $\xi$, one trend parameter $\beta$, and two swim-suit parameters $\gamma_1$ and $\gamma_2$ for each of the 34 events, giving a total of 204 independent parameters. These parameters, after the transformation described below, are plotted against $u_{L,e} = \log(-u_e)$, recalling that the data are negative, with $u_{e} < 0$, so $u_{L,e}$ is the log of the 200th best swim-time for event $e$ in the data. For each of the transformed parameters $$\sigma_{L}^{(e)} = \log(\tilde{\sigma}^{(e)}_u), \;\mu_L^{(e)} = \log\left(-\mu_0^{(e)}\right), \;\beta_L^{(e)}=\log\left(\beta^{(e)}\right), \;\gamma_{L,1}^{(e)} = \sqrt{\gamma_1^{(e)}}, \;\gamma_{L,2}^{(e)} = \sqrt{\gamma_2^{(e)}},$$ there is some linear or near-linear relationship with $u_{L,e}$, and $\xi^{(e)}$ is approximately constant. In the case of the location parameter $\mu_0^{(e)}$, this is a consequence of the choice of threshold. More generally, power law relationships are commonly found in sports \citep{sylvan1999power}, and the connection between $u_e$ and $\tilde{\sigma}^{(e)}_u, \mu^{(e)}_0, \beta^{(e)}$ was hypothesised based on the prevalence of log-log relationships in sports modelling \citep{riegel1981athletic}. This relationship however, does not explain the dependence between swim-time and swim-suit effects $\gamma_1^{(e)}$ and $\gamma_2^{(e)}$ well, which is a combined result of the biomechanical and physical relationship between range of movement and flexibility, drag, buoyancy and total energy expenditure amongst other factors. The reason for this complex relationship is not explored in this article, but was chosen based on a Box-Cox transformation in the single suit case of $\gamma$ to $\gamma^*$, such that 
$$\gamma^* = 
\left\{ {\begin{array}{lc}
(\gamma^{\delta_\gamma} -1)/\delta_\gamma & \delta_\gamma \neq 0, \\
\log(\gamma) & \delta_\gamma  =0, \\
 \end{array} } \right.$$
where $\gamma^*$ is assumed to come from a model which is linear in $u_{L,e}$ with a normal error distribution with constant variance. The choice of $\delta_\gamma~=~1/2$ is consistent with the Box-Cox transformation, which gives an MLE and 95\% confidence interval of $\delta_\gamma = 0.52 \; (0.37,0.68)$, and also agreed with Box-Cox transformation applied to $\gamma_1^{(e)}$ and $\gamma_2^{(e)}$ in the two-suit case.  Box-Cox transformations were also applied to the other parameters to confirm the log-log hypothesis, for example $\delta_\beta =  0.049 \; (-0.11,  0.19)$, indicating that a  log relationship is appropriate.
These relationships motivate the across event model of the next section.

\subsection{Across Event Model}

\subsubsection{Parametric Model\label{sec:paraModel}}

Now that models \eqref{eq:parameterisation} and \eqref{eq:parameterisation_twosuit} have been shown to be suitable for each event, it is desired that information can be shared between events to reduce parameter uncertainty and improve predictive performance. By doing this we ensure that the across event model is more robust than models \eqref{eq:parameterisation} and \eqref{eq:parameterisation_twosuit} with respect to anomalous data, which could lead to over-fitting.

A natural first step here would be to consider distance as a covariate and a log-log relationship. Distance does work well in athletics, as long as it is within the same gender \citep{riegel1981athletic}. However, distance does not work well when pooling across both genders, and across different strokes, since for example breaststroke is always slower than freestyle for a given distance and gender, and so inherent bias will be introduced due to the physical nature of the difference in strokes. Instead, the threshold swim-time is used as a covariate, since naturally slower strokes, whose corresponding scale parameters for example are likely to be larger, will also have a larger covariate, the threshold swim-time. This allows for a given parameter to vary smoothly across events, rather than to be discretised by the distance of the event. Thus, no adjustment is needed to compare between different strokes and genders.

From Figure \ref{fig:paramplots} it is initially hypothesised that the shape parameter $\xi^{(e)}$ can be held constant across all events, and that the transformed parameters $\sigma_{L}^{(e)}$, $\mu_L^{(e)}$, $\beta_L^{(e)}$, $\gamma_{L_1}^{(e)}$ and $\gamma_{L_2}^{(e)}$ increase linearly with $u_{L,e}$. A similar figure (not shown) exists for the single-suit parametrisation, which suggests linearity for $\gamma_{L}^{(e)}$ also.
Thus, it is proposed that the parameters are pooled across the 34 events via the following model:
\begin{eqnarray}
\label{eq:Model3.0}
\xi^{(e)} & =& \xi, \\
\label{eq:Model3.1}
\mu_L^{(e)} &=& \alpha_1 + \vartheta_1 u_{L,e},\\
\label{eq:Model3.2} 
\sigma_{L}^{(e)}& =& \alpha_2 + \vartheta_2 u_{L,e},\\
\label{eq:Model3.3}
\beta_L^{(e)} &=& \alpha_3 + \vartheta_3 u_{L,e}.
\end{eqnarray} In the single-suit case,
\begin{equation}
\label{eq:Model3.5a}
\gamma_L^{(e)} = \alpha_4 + \vartheta_4 u_{L,e},
\end{equation}
and in the two-suit case,
\begin{equation}
\label{eq:Model3.5b}
\gamma_{L_1}^{(e)} = \alpha_4 + \vartheta_4 u_{L,e},\;
\gamma_{L_2}^{(e)} = \alpha_4 + \varepsilon + \vartheta_4 u_{L,e},
\end{equation}
for some parameters $\pmb{\psi} = \{\xi,\varepsilon, \; \{\alpha_i, \vartheta_i \in \mathbb{R}: \;  i = 1,\dots,4 \} \}$. Having two separate gradients, $\vartheta_4$ and $\vartheta_5$ such that $ \gamma_{L_1}^{(e)} = \alpha_4 + \vartheta_4 u_{L,e},$ and $
\gamma_{L_2}^{(e)} = \alpha_4 + \varepsilon + \vartheta_5 u_{L,e}$, was also considered, but it was found that a common gradient, such that $\vartheta_5 = \vartheta_4$, sufficed. In fact, several other models were considered (not reported), for example including a different intercept for men's and women's events in the linear model, or using separate linear models for different distances, but these were found to produce no improvement. The full likelihood of the across event parametric model then, assuming independence between events, is therefore given as 
\begin{equation*}
L(\pmb{\psi}; \pmb{x}) = \prod_{e \in E}\left\{\exp\left[-\Lambda^{(e)}\left(\mathcal{A}_{1,u}\right)\right] 
\prod_{i=1}^{200} \int_{x_i^{(e)}-s/2}^{x_i^{(e)}+s/2}\lambda^{(e)}(t_i^{(e)},x) \diff x\right\}.
\end{equation*}

\begin{table}[h!]
\centering
\begin{tabular}{@{\extracolsep{5pt}}clll}
\\[-1.8ex]\hline 
\hline \\[-1.8ex] 
Model& Constraints & AIC/RIC &  \# ind. parameters \\
\hline \\[-1.8ex]
$\mathcal{M}_{1a}$ &independent fits, single-suit \eqref{eq:parameterisation}& $0$ & $170$ \\
$\mathcal{M}_{1b}$ &independent fits, two-suits \eqref{eq:parameterisation_twosuit}& $-23.7$ & $204$ \\
$\mathcal{M}_2$ & $\mathcal{M}_{1a}$ with constraint \eqref{eq:Model3.0} &$-38.7$ & $137$ \\
$\mathcal{M}_3$ & $\mathcal{M}_2$ with constraint \eqref{eq:Model3.1} &$ -52.1$ &$ 105$ \\
$\mathcal{M}_4$ & $\mathcal{M}_3$ with constraint \eqref{eq:Model3.2} & $-29.6$ &$73$ \\
$\mathcal{M}_5$ & $\mathcal{M}_3$ with constraint \eqref{eq:Model2} & $-58.7$ & $74.1$ \\
$\mathcal{M}_6$ & $\mathcal{M}_5$ with constraint \eqref{eq:Model3.3} & $-87.7$ & $42.1$ \\
$\mathcal{M}_{7a}$ & $\mathcal{M}_6$ with constraint \eqref{eq:Model3.5a} & $-90.6$ & $10.3$ \\
$\mathcal{M}_{7b}$ & $\mathcal{M}_{1b}$ with constraints \eqref{eq:Model3.0}, \eqref{eq:Model3.1},  \eqref{eq:Model3.3}, \eqref{eq:Model3.5b}, \eqref{eq:Model2} & $-121.5$ & $11.2$\\
\hline \\[-1.8ex] 
\end{tabular}
\caption{Model comparison showing the AIC or RIC for each model, normalised by the independent fits model with a single suit, model $\mathcal{M}_{1a}$. The RIC, defined by expression \eqref{eq:RIC}, is used when a spline is fitted to a parameter over events and defines the number of effective degrees of freedom. A lower AIC or RIC indicates a better model fit.
}
\label{tab:model_comaparison}
\end{table}
This pooled structure was incrementally implemented as shown in Table \ref{tab:model_comaparison}. The first model fitted, $\mathcal{M}_{1a}$ pools no parameters and considers only a single suit, such that each event $e$ has 5 independent parameters, $(\mu^{(e)}_0, \sigma^{(e)}_u, \xi^{(e)}, \beta^{(e)}, \gamma^{(e)})$, resulting in a total of 170 parameters. The AIC can be seen to improve from $\mathcal{M}_{1a}$ to $\mathcal{M}_{1b}$ by including the separate effect of two suits, despite the significant increase in the number of free parameters. From model $\mathcal{M}_{1a}$, the pooling structure begins to be implemented, and there is an improvement to $\mathcal{M}_2$, where now constraint \eqref{eq:Model3.0} is introduced such that all events share a common shape parameter. Again, the model fit improves from $\mathcal{M}_2$ to $\mathcal{M}_3$ by employing constraint \eqref{eq:Model3.1}, however, when trying to enforce linearity between $\sigma_{L}^{(e)}$ and $u_{L,e}$ across $e \in E$ via constraint \eqref{eq:Model3.2}, model $\mathcal{M}_4$, the fit was poorer. The events which mainly contributed to this worsened fit were the men's and women's 200m free and women's 50m fly, but the fit was also generally worse across the vast majority of events, which could be explained by some non-linearity observed in Figure \ref{fig:paramplots}.

The inadequacy of a linear relationship \eqref{eq:Model3.2} between $\sigma_{L}$ and $u_{L}$ suggests that a fully parametric model to describe this relationship was slightly too restrictive, and motivates the need for a more flexible but parsimonious model, for which we use semi-parametric techniques. Model $\mathcal{M}_5$ was therefore introduced which relaxes the linear constraint \eqref{eq:Model3.2} on $\sigma_{L}$, and instead uses the spline based non-parametric approach described in Section \ref{sec:semi_par}, which lets the smooth dependence of $\sigma_{L}$ on $u_{L}$ to be captured by allowing the data to govern the precise nature of this relationship, whilst keeping the dependencies of $\mu$ and $\xi$ on $u_L$ the same and keeping $\beta$ and $\gamma$ unconstrained, as in model $\mathcal{M}_4$. From here, models $\mathcal{M}_6$ and $\mathcal{M}_{7a}$ are then fitted by cumulatively employing constraints \eqref{eq:Model3.3} and \eqref{eq:Model3.5a} respectively, and finally $\mathcal{M}_{7b}$ is fitted by the addition of an extra suit parameter to $\mathcal{M}_{7a}$
, see Table \ref{tab:model_comaparison}. The best fitting model, determined via regularisation information criteria (RIC) \citep{shibata1989statistical} which is defined by expression \eqref{eq:RIC}, is $\mathcal{M}_{7b}$ with only approximately 11 parameters. Critically, note the substantial improvement from models $\mathcal{M}_{7a}$ to $\mathcal{M}_{7b}$, showing a clear impact of changes in full body suit technology over the period when these suits were allowed.

Confidence intervals were found via parametric bootstrapping, such that model $\mathcal{M}_{7b}$ was re-fitted to 250 simulated datasets, to estimate the sampling distribution of parameter estimators. The number of observations from event $e$ in simulated dataset $j$, $N^{(e)}_j$ is simulated directly via, $N^{(e)}_j \sim \textup{Poisson} \left(\Lambda^{(e)}\left(\mathcal{A}_{1,u} \right)\right) $. For an event $e$ and replication $j$, $N^{(e)}_j$ swim-times $x^{(e)}_1, \dots, x^{(e)}_{N^{(e)}_j}$ and the time of these swims $t^{(e)}_1, \dots, t^{(e)}_{N^{(e)}_j}$ were generated via a probability integral transform on equation \eqref{eq:GPd} for the swim-times, and the distribution function \eqref{eq:intensityDensT} integrated over $x$ for the times respectively. Some of the resulting bootstrapped parameter estimates resulted in infeasible estimates, for example inferring that the ultimate possible swim-time is worse than some swim-times in the original data set, or that the expected next world record swim-time is worse than the current world record, and so these data sets were discarded. The remaining 240 data sets quantify the natural variation in the data and thus provide the basis for obtaining confidence intervals. All confidence intervals referred to subsequently in this article are obtained via this method.
 	
The estimated values for $\vartheta_3$ and $\vartheta_4$ under model $\mathcal{M}_{7b}$, the associated gradients for the trend parameters and swim-suit parameters respectively, were $\hat{\vartheta}_3 = 0.940\; (0.936,0.942)$ and $\hat{\vartheta}_4 =0.460 \; (0.432, 0.470)$. The relative confidence interval widths are smaller on $\vartheta_3$ than $\vartheta_4$, and this is likely due to the swim-suit parameter being dependent on less data than the trend parameter, since only data in swim-suit years effect it. In comparison, the gradient governing the linear relationship \eqref{eq:Model3.1} is estimated at $\hat{\vartheta}_1 = 1.0016\; (1.0010 , 1.0019)$. The tight confidence intervals here indicate the strong relationship between $u_L$ and $\mu_L$.
\subsubsection{Semi-Parametric model}
\label{sec:semi_par}


To achieve the appropriate flexibility to model the relationship observed in Figure \ref{fig:paramplots} between $\sigma_L$ and $u_L$ we use a $d$-degree spline function \citep{de1978practical}, which is a piecewise polynomial function that is constructed to be continuous and $d$ times continuously differentiable over a closed interval domain. It is a weighted linear sum of $q$, $d$-degree basis splines, called B-splines, with the $k^{th}$ B-spline $B_k(x)$ centred on a \textit{knot} at point $x_k$. The spline function used for $\sigma_L$ is denoted by \begin{equation}
\label{eq:Model2}
\sigma_{L}(u_{L}) = \sum_{k=1}^q a_k B_{k}(u_{L})
\end{equation} where $a_k$ is the $k^{th}$ element of the spline coefficient vector $\pmb{a} = (a_1,\dots,a_q)$ which is constant over all events such that, given a vector $\pmb{a}$, the value of $\sigma_L$ for any given event $e$ is a function of $u_{L}$ only, see Appendix \ref{sec:appendSpline} for further details. 

Although function \eqref{eq:Model2} can model any non-linear relationship, we wish for this relationship to be smooth and increasing. In order to enforce this smoothness, the likelihood function is extended to a penalised likelihood which contains a roughness penalty.  The penalty is governed by $\phi_r p_r = \phi_r \pmb{a}^T P \pmb{a}$, where $P \in \mathbb{R}^{q\times q}$  is the penalty matrix, and $\phi_r > 0$ determines the amount of penalisation. The choice of $P$ determines the nature of the penalty and is chosen based on the form of the data, or some prior belief. In this case a 2nd order penalty on the finite differences of adjacent coefficients \citep{eilers1996flexible}, and a degree $d=4$ spline was chosen, see Appendix \ref{sec:appendSpline}. This penalises $\sigma_L$ having a large second derivative, and penalises fits for $\sigma_L$ that depart from linearity. Additionally, since it is believed apriori that the GPd scale parameter is an increasing function of the threshold swim-time, a hard constraint $\phi_m p_m$ ensures monotonicity in the spline function, where $p_m$ is defined as follows: allow $$\{z_1, \dots, z_k\} = \left\{\min_{e \in E} u_{L,e} \right\} \cup   \left\{x_i: \frac{\diff\sigma_L(x_i)}{\diff x}=0, \; i=2,\dots,k-1\right\} \cup \left\{\max_{e \in E} u_{L,e} \right\}$$ to be a discrete set of size $k$ containing all stationary points and end points of the spline function, then $$p_m = -\sum_{i=1}^{k-1} \left(\sigma_L(z_{i+1}) - \sigma_L(z_{i}) \right) \mathds{1}\left\{\sigma_L(z_{i+1})-\sigma_L(z_{i}) < 0  \right\}.$$

With the GPd scale parameter $\tilde{\sigma}_u$ for a particular event $e$ being defined by the spline via
\begin{equation*}
\tilde{\sigma}_u^{(e)} = \exp \left[ \sum_{k=1}^{q} a_k B_k(u_{L,e}) \right],
\end{equation*}
the full joint penalised likelihood across all events becomes 
\begin{align*}
L_p(\pmb{\varphi}, \phi_r, \phi_m; \pmb{x}) &= \prod_{e\in E}\left\{\exp\left[-\Lambda^{(e)}\left(\mathcal{A}_{1,u}\right)\right] 
\prod_{i=1}^{200} \int_{x_i^{(e)}-s/2}^{x_i^{(e)}+s/2}\lambda^{(e)}(t_i,x) \diff x \right\}\exp \left[ -(\phi_r p_r + \phi_m p_m)\right],\\
&=L(\pmb{\varphi}; \pmb{x}) \exp\left[ -(\phi_r p_r + \phi_m p_m)\right],
\end{align*}
where $\pmb{\varphi}$ are the parameters of the model, and $L$ is the unpenalised likelihood. The penalised log-likelihood for model $\mathcal{M}$ is therefore given as $$\ell_p(\mathcal{M}) = \ell (\mathcal{M}) - \phi_r p_r - \phi_m p_m,$$ where $\ell$ is the unpenalised log-likelihood, and $\phi_m > 0$ is sufficiently large such that monotonicity is a hard constraint. The value of $\phi_m$ is found  by finding a $\phi_m$ such that $$\max\left(\ell_p(\mathcal{M}|\phi_m)\right) = \max\left(\ell_p(\mathcal{M}|\phi_m+\epsilon)\right),$$ for any $\epsilon>0$. Theoretically, this can be found by allowing $\phi_m \rightarrow \infty$, however it can be difficult for optimisation routines to converge to this global maxima. Therefore, in practise $\phi_m$ is increased iteratively by initially setting $\phi_m=0$ and finding the parameter that give $\max\left(\ell_p(\mathcal{M}|\phi_m=0)\right)$. Then $\phi_m$ is increased iteratively, using the previous solution as the initial starting parameters, until there is no change in $\mathcal{M}$ and therefore also no change in $\ell(\mathcal{M})$. Instead of a constraint on the spline function itself to enforce monotonicity, I-splines \citep{ramsay1988monotone} could have been used as a basis instead of B-splines, and then positivity constraints on the basis splines would have enforced monotonicity. This construction may have resulted in more efficient computation, but would yield essentially identical model fits and results.

\begin{figure}
\centering
\includegraphics[scale=0.7]{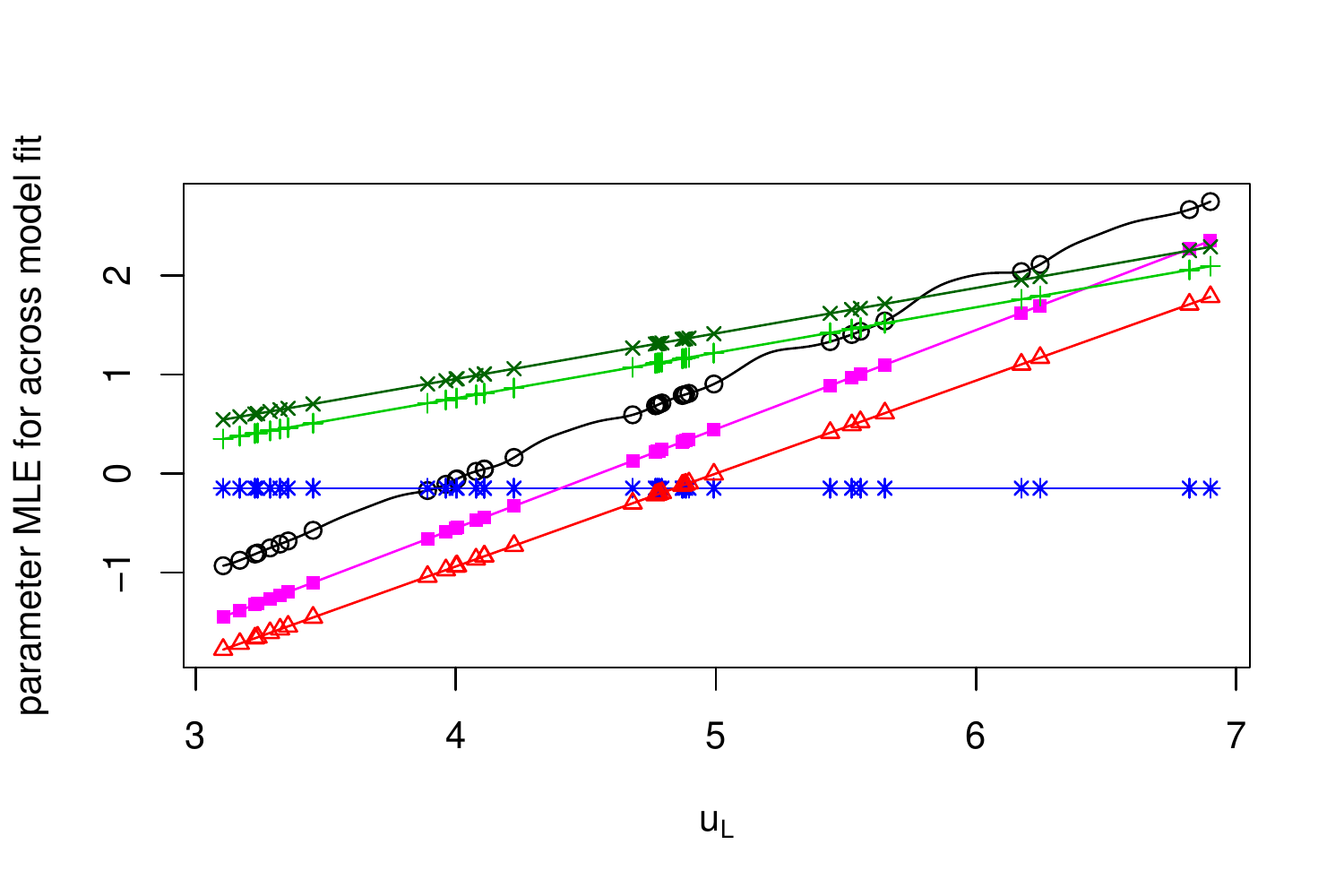}
\caption{Fitted parameters for model $\mathcal{M}_{7b}$, as a function of $u_L$: $\sigma_{L}(u_L)$ (black circles) is governed by the spline, whilst $\beta_L(u_L)$ (red triangles), $\mu_L(u_L)-5$ (purple squares), $\gamma_{L_1}(u_L)$ light green pluses, $+$) and $\gamma_{L_2}(u_L)$ (dark green crosses, $\times$) vary linearly with $u_L$. The shape parameter (blue stars) has a constant value of $\hat{\xi} = -0.147\;  (-0.152, -0.143)$. Note that $\mu_L$ has been rescaled, by subtracting $5$ uniformly, to be visible on the plot.}
\label{fig:paramplots_Semi}
\end{figure}

The choice of $\phi_r$ is selected using 10-fold cross validation to maximise model predictive performance at data points not used for fitting \citep{ewans2008effect}. The model is fitted based on a random stratified sample of 90\% of the data, the training data, which is then used calculate the log-likelihood based on the remaining 10\% of the data, the test data. The log-likelihood for each of the 10 non-overlapping sets of test-data is summed to obtain a `predictive' log-likelihood based on the prediction accuracy of the model. This process is repeated 20 times at a range of different values of $\phi_r$, with the value of $\phi_r$ which corresponds to the best average predictive performance being selected as the optimum penalty. It was found that the change in predictive log-likelihood was robust to changes in $\phi_r$, and it is thought that this is due to the hard constraint on monotonicity already accounting for much of the variability in the spline fits. For model $\mathcal{M}_{7b}$, an optimum penalty of $\phi_r= 15$ was found. Given this, the full model can be fitted and the parameters as a function of $u_L$ are shown in Figure \ref{fig:paramplots_Semi}.

Since models $\mathcal{M}_5$, $\mathcal{M}_6$, $\mathcal{M}_{7a}$ and $\mathcal{M}_{7b}$ are semi-parametric, AIC can no longer be used as a model comparison tool since the number of degrees of freedom is not defined. Instead, RIC is used, which uses the effective degrees of freedom $g$, as opposed to degrees of freedom. Otherwise, RIC is defined identically to AIC, that is \begin{equation}
\label{eq:RIC}
\textup{RIC} = -2 \ell(\pmb{\varphi}) + 2 \; \textup{tr}\left[ I(\pmb{\varphi}) J(\pmb{\varphi}, \phi_r, \phi_m)^{-1}\right],
\end{equation} such that $g=\textup{tr}\left[ I(\pmb{\varphi}) J(\pmb{\varphi}, \phi_r, \phi_m)^{-1}\right]$ where $I$ is the observed Fisher information criteria of the unpenalised likelihood $L$, $J$ is the negative Hessian matrix of the penalised log-likelihood $L_p$, and $\textup{tr}(A)$ is the trace of the square matrix $A$.
\subsubsection{Assessment of model $\mathcal{M}_{7b}$ fit}
\begin{figure}[!htbp]
\centering
\begin{minipage}{.45\textwidth}
  \centering
  \includegraphics[scale=.5]{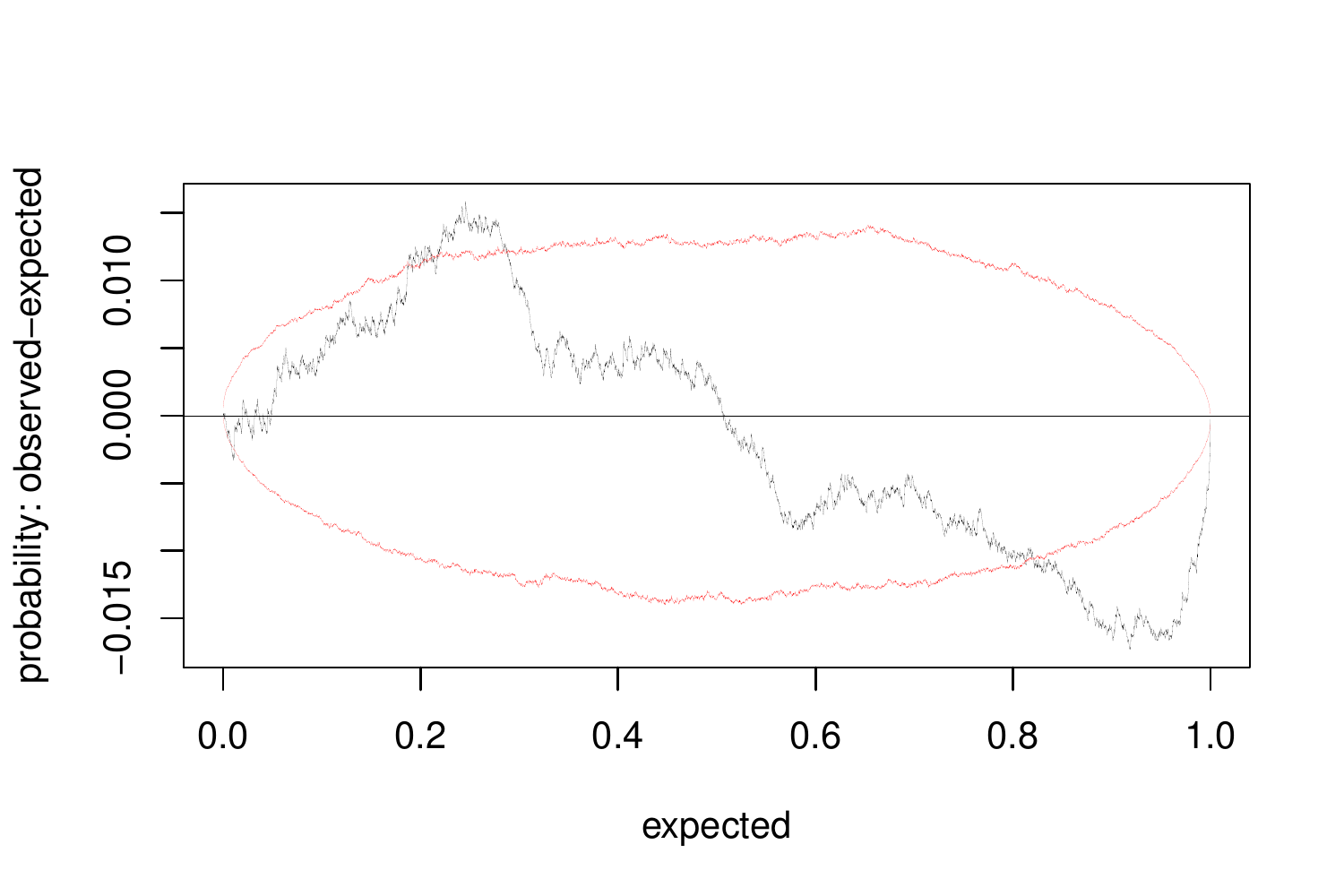}

\end{minipage}%
\begin{minipage}{.05\textwidth}
\hspace{1mm}
\end{minipage}
\begin{minipage}{.45\textwidth}
  \centering
  \includegraphics[scale=.5]{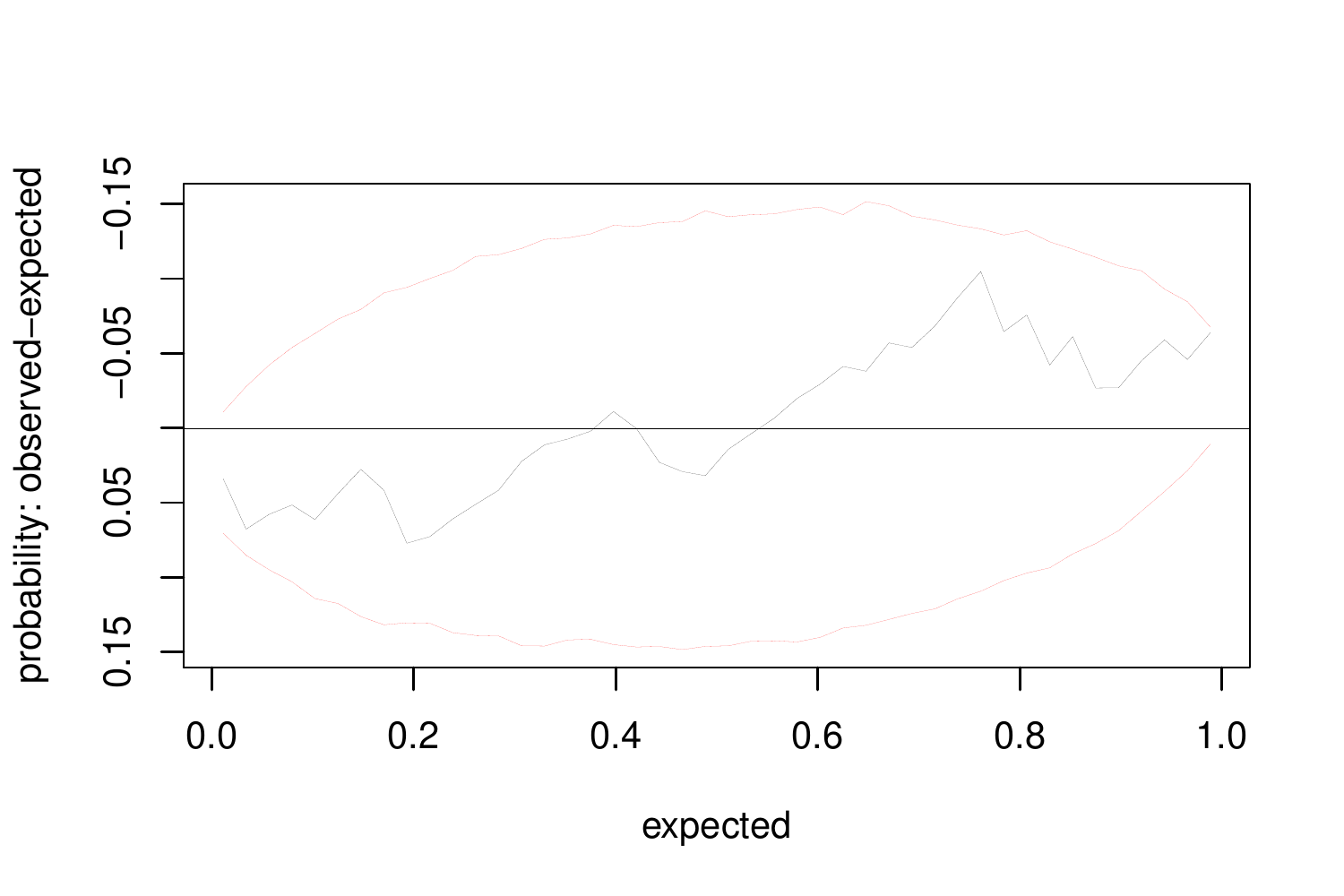}

\end{minipage}
\caption{PP plot (plotted as observed minus expected probabilities) pooled over all events, with 95\% tolerance intervals, using both the whole data set (left) and only data from $[2001,2003]$ (right).} 
\label{fig:PPplot}
\end{figure}

The rate of exceedances and the distribution above threshold must both be considered to determine the overall quality of the selected model fit. A pooled PP plot is used to determine how well the model fits the distribution of swim-times above threshold. The pooled PP plot, Figure~\ref{fig:PPplot} (left), allows the combined fit of all 34 events to be analysed at once. The fit generally is very good, especially considering the reduction from 204 to 11.2 parameters. The areas of weaker fit can mainly be attributed to two events, the 200m men's free, and the 50m men's fly. These two events increase the RIC by 10.4 and 9.7 respectively, both of which is significant evidence of lack of fit, so caution should be exercised when drawing conclusions from these two events. Somewhat surprisingly though, we find that removing these events from the analysis makes no substantial difference to the diagnostic shown in Figure \ref{fig:PPplot} (left). Figure \ref{fig:PPplot} (right) shows another pooled PP plot, using the same model fit, but only using data from the period $[2001,2003]$. These data also appear to be fit very well, and this implies that
any potential bias introduced by the early period data selection problems, highlighted in Section \ref{sec:thedata},
is minimal.

A nice feature of this pooled model is that natural ordering across different strokes is preserved even for events which carry a less good fit. For example, the parameters for the 50m men's fly will always indicate that it is a faster event than the 50m men's breaststroke, i.e., by predicting a faster ultimate possible swim-time or next world record swim-time.

\begin{figure}[!htpb]
\centering
\includegraphics[scale=0.7]{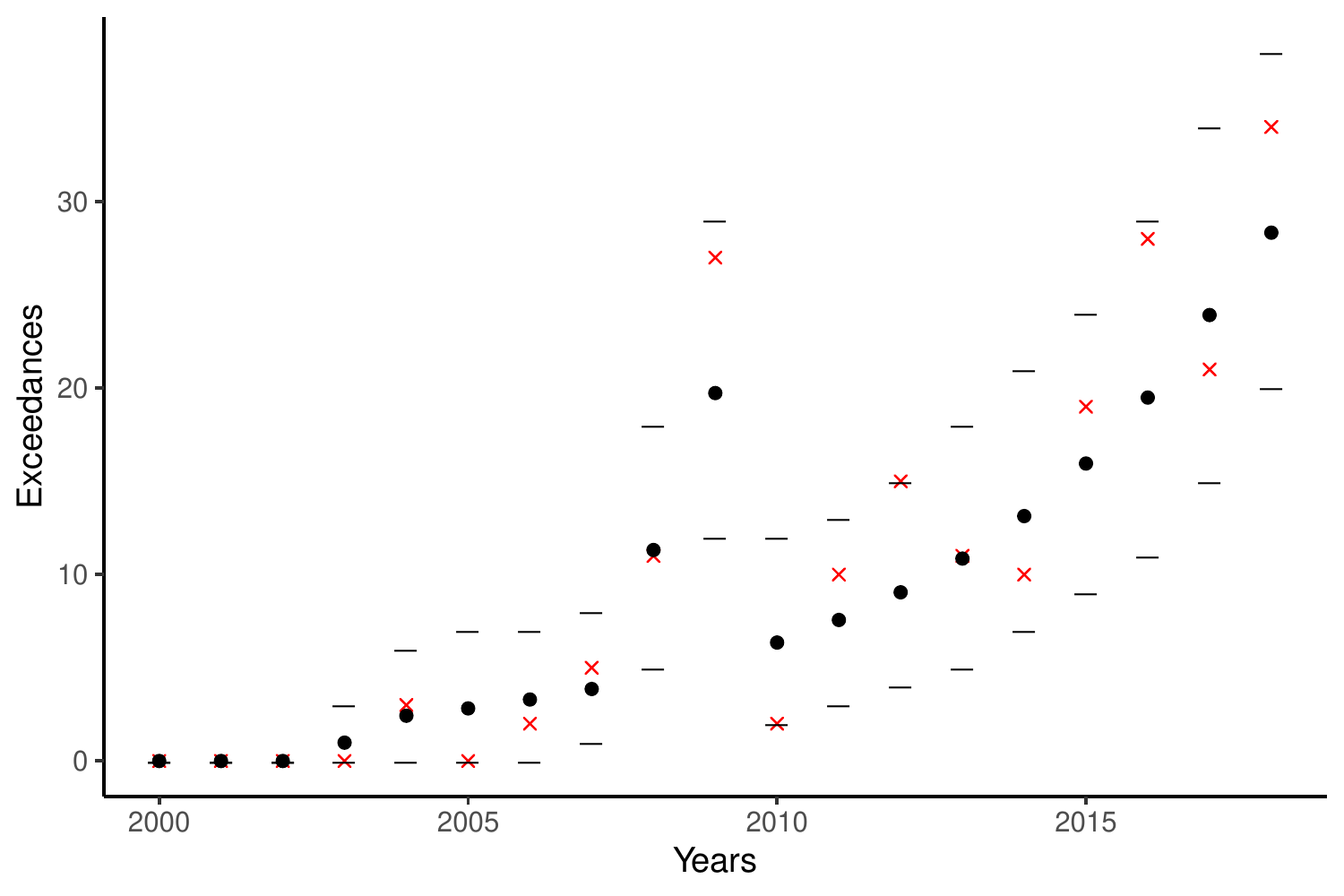}
\caption{The estimated expected (black circles) and observed (red crosses) number of observations per year better than $u_e$ for women's 100m freestyle, with 95\% confidence intervals for the estimated values given by the lower and upper horizontal lines. The two swim-suit years, 2008 and 2009, have increased rates of exceedances relative to neighbouring years.}
\label{fig:rate_check}
\end{figure}

Figure \ref{fig:rate_check} shows the expected rate of observations exceeding $u_e$ per year, compared to what was observed in the data, for the women's 100m freestyle. Similar plots for all 34 events were examined (not shown). It can be seen that the observed rate of observations almost always falls
between (and once only marginally outside) the 95\% confidence intervals, including during the swim-suit era and the early period of the database when competition selection may have induced bias as identified in Section \ref{sec:thedata}. The estimated expected number of observations is not systematically above or below the observed number of observations. For a year in which the observed rate is higher than expected, often in the next year this observed rate is below the expected rate, which is due to the discrete nature of the plot.

%

\section{Results from Model \label{sec:results}}

\subsection{Rankings}

From fitting model $\mathcal{M}_{7b}$, the final rankings of the best ever swim-times can be constructed. The rankings are determined by the $r$-value of a swim-time $x$, that is, the rate at which observations better than $x$ occur in the given event. If $X^{(e)}_{t}$ is the random variable denoting a new observed negative swim-time in event $e$ at a time $t$ where this swim-time is better than $u_e$, then the expected rate $R$ at which an observation $X_t^{(e)}$ is faster than swim-time $x$ occurs is defined as follows:
\begin{align}
\label{eq:rankings_}
R\{X_t^{(e)} > x+s/2\} &= \Pr\{X_t^{(e)} > x+s/2 | X_t^{(e)} > u_e\} \Lambda^{(e)}_{y(t)}\left(\mathcal{A}_{1,u}\right) \nonumber\\
&= \bar{H}^{(e)}_{u}(x+s/2)\Lambda^{(e)}_{y(t)}\left(\mathcal{A}_{1,u}\right) \nonumber\\
 &\approx  \left[1+\xi\left(\frac{x+s/2-u_e}{\tilde{\sigma}^{(e)}_{u}}\right)\right]_+^{-\frac{1}{\xi}} 
\left[1+\xi \left(\frac{u_e-\mu^{(e)}(y^*(t))}{\sigma^{(e)}(y^*(t))}\right)\right]^{-\frac{1}{\xi}}_+ ,
\end{align}
for all $x+s/2>u_e,$ where the final approximation follows from equation \eqref{eq:rate_approx}, where $y(t)$ is the year in which $X_t^{(e)}$ occurs and $y^*(t)=y(t)+1/2$ is the mid point of years $y(t)$ and $y(t) +1$. An estimate of $R\{X_t^{(e)} > x+s/2\}$ gives the $r$-value, and therefore a measure of the `quality' of the swim-time $x$. By adding $s/2$, the censoring is taken into consideration, since the true observed swim-time $X_t^{(e)}$ would need to be faster by an amount greater than the precision of the data to be recorded as being faster. 

\begin{figure}[h!]
\centering
\includegraphics[trim={.4cm 0 0 0},clip, scale = 0.8]{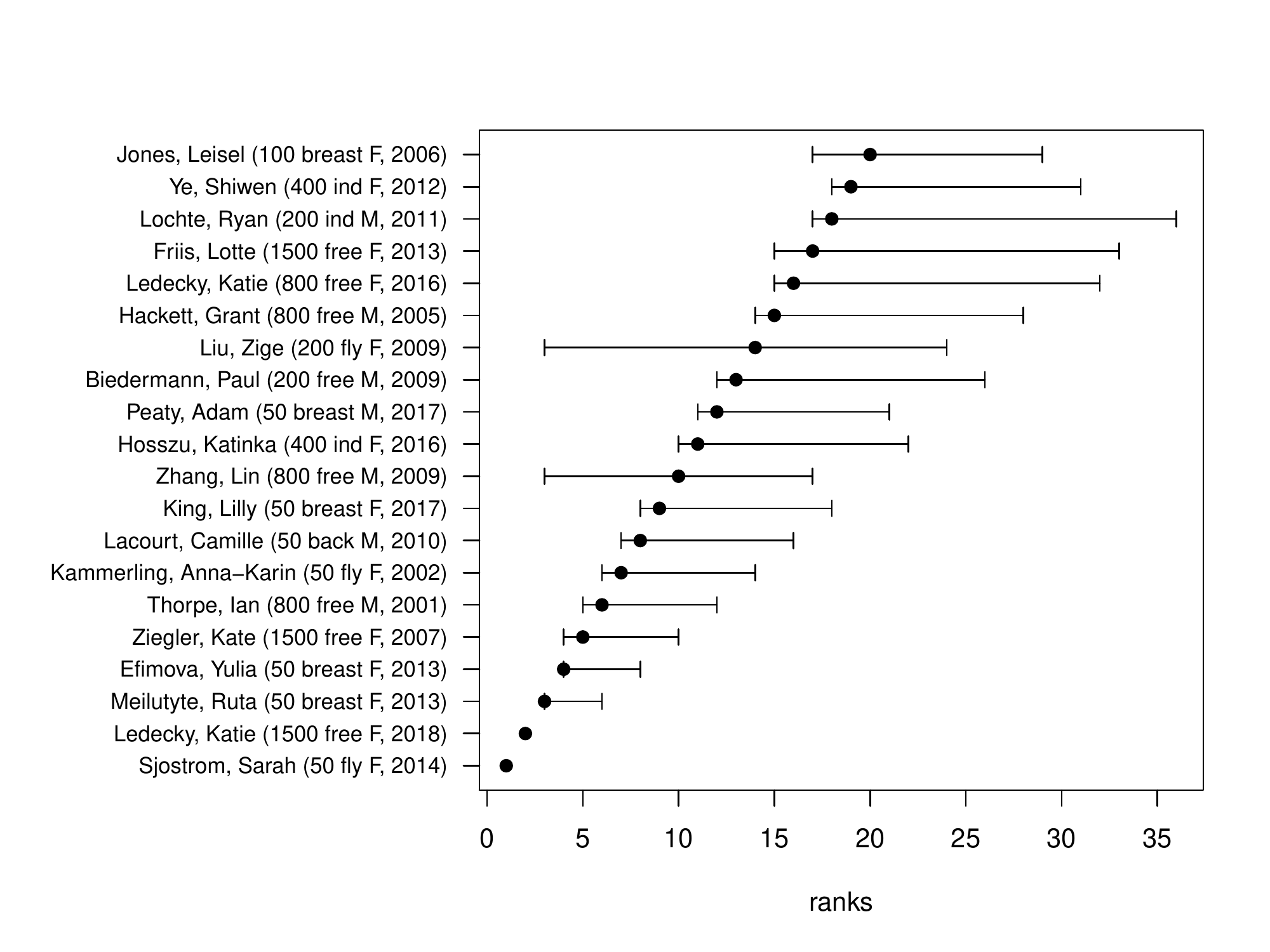}
\caption{The ranking of the top 20 swimmers from the data set, with 95 \% CIs from bootstrapped data sets. Better ranked swimmers are lower on the y-axis.} 
\label{fig:ranks_plot}
\end{figure}

Figure \ref{fig:ranks_plot} shows the best 20 swimmers from the 2001 to end of 2018 period, based on the $r$-value of their swim. Note that swimmers names can occur multiple times where they have recorded swim-times in more than one event. The error bars show the $95$\% confidence intervals from the parametric bootstrapping. It is also possible to quantify how much better one swimmer is than another by analysing what proportion of time the bootstrapped samples give one swimmer ranked ahead of another. For example, Adam Peaty, ranked 12th, beats Katinka Hosszu, ranked 11th, on $48$\% of rankings from the bootstrapped data sets. In contrast, Katie Ledecky's 1500m free performance, ranked 2nd, never beats top ranked Sarah Sjostrom's 50m fly performance, giving strong evidence for ranking Sarah Sjostrom better. 

The lower confidence intervals for the ranks of both Zige Liu and Lin Zhang are much wider in comparison to the others in the top 20, and one possible reason is that they were swam during the second swim-suit period, $S_{t_2}$ (2009). As noted in Section \ref{sec:paraModel}, the relative uncertainty for $\vartheta_4$, which controls the swim-suit effect, is comparatively large, and this added uncertainty propagates through to the rankings. Essentially, the confidence intervals are showing that, if the parameter associated with the 2009 swim-suit is overvaluing the effect of a this suit, then their true ranks could be much lower. This same effect is not seen in Paul Biedermann's rank, also swam in 2009, however this was in the 200m men's free which has previously been identified as an area of weaker fit.
  
Interestingly, in some cases the time when the swim was performed can effect the order of the rank within the same event. Ruta Meilutyte, Yulia Efimova and Lilly King hold ranks 3, 4, and 9 respectively, all from the 50m women's breaststroke, however the fastest time of the three is Lilly King's with a time of 29.40 seconds in July 2017, compared to times of 29.48 seconds and 29.52 seconds for Ruta Meilutyte and Yulia Efimova respectively, which were both swam in July 2013, five years earlier, which indicates that they achieved comparatively better results given their era. 

It is worth noting that 7 of the top 20 estimated ranked swims occur in 50m races, which is approximately 50\% more than the number that would be expected if the assumption that all events are equally competitive holds. In fact, this assumption is unlikely to hold in practice, since the 50m backstroke, breaststroke and fly are non-Olympic events, and as such the competitiveness of these events may be less than the Olympic events, which increases the disparity between the observed and expected number of 50m races in the top 20 ranks. Conversely, the top 20 rankings for the independent fits model $\mathcal{M}_{1a}$ and $\mathcal{M}_{1b}$ (not shown), were found to be proportionately represented by all distances. In models $\mathcal{M}_{1a}$ and $\mathcal{M}_{1b}$ fits the 50m events have larger corresponding shape parameters than other events on average, and than the common shape parameter for $\mathcal{M}_{7b}$, particularly the men's fly and women's and men's free had comparatively much larger shape parameters than other events. Therefore, it was initially thought that high rankings of swimmers in the 50m events may be due to the enforcing of a constant shape parameter across all events, and so perhaps a different modelling strategy is required for the shorter events. However, it was found from calculating profile likelihood based 95\% confidence intervals that the the shape parameters were $-0.067\; (-0.221,-0.045)$, $0.000 \; (-0.173,0.013)$ and $-0.080\; (-0.253,-0.090)$ for the men's fly and women's and men's free respectively, which all overlap with the shared shape parameter of model $\mathcal{M}_{7b}$, $\hat{\xi} = -0.147\;  (-0.152, -0.143)$. Thus, it appears that the comparatively larger shape parameters for the 50m events is mostly due to natural variation. A more formal test for a different shape parameter, common to all 50m events, would be to evaluate the RIC under a model with indicator covariates for the shape parameters for these 50m events. This was not considered necessary given the evidence from the profile likelihood intervals given above. Notably, the ordering of the very top four ranks was the same under both $\mathcal{M}_{1b}$ and $\mathcal{M}_{7b}$.

A national rankings table can also be made by only including a given nation in the comparison, and could be used for that nation's Olympics selection, for example. This would also change the confidence intervals for the rankings as swimmer's are only compared to others from the same nation.

\subsection{Ultimate times}
Finding limits to human sports performance has interested academics for years, in athletics for example \citet{blest1996lower}. In swimming, \cite{nevill2007there} attempt to determine the ultimate possible time by analysing world record swims from 1957 to 2007, and \cite{toussaint122005biomechanical} approach this from a biomechanical perspective. In this article, the ultimate time is determined from the GPd function.

It was found that the MLE for the shape parameter with 95\% confidence intervals was $\hat{\xi} = -0.147\;  (-0.152, -0.143)$ which (since $\hat{\xi} < 0 $) can be interpreted as there being a finite bound on the best possible time a human can achieve in any given event. In many applications getting such a narrow confidence intervals, and hence such clear evidence $\xi<0$, is difficult to achieve. Here this has been enabled by the pooling of data from all 34 events, giving a sample from the model of 6800 observations to inform us of the value of $\xi$. The ultimate possible time for an event can be estimated directly from the parameter estimates since for event $e$ there exists an end-point $x_{H,e} = u_e - \tilde{\sigma}^{(e)}_u/\xi: H^{(e)}_u(x)=1,\; \forall \; x>x_{H,e}$. Note that $x_{H,e}$ is covariate independent in the selected model, which seems reasonable since we expect the gap between the ultimate possible time and the world record to shrink as world records improve, but the ultimate time is still unreachable and `set-in-stone'. For example, the MLE for the ultimate possible time for the men's 100m breaststroke given by the model is $53.81 \; (53.60, 53.97)$ seconds. In comparison, Adam Peaty's fastest time in the dataset is $57.10$ seconds from 2018. This 3 second difference made Peaty's \textit{Project 56} (\url{https://www.bbc.co.uk/sport/av/swimming/40650276}), his challenge to swim a sub 57s 100m breaststroke, seem more achievable than at first glance. In fact, Peaty has since succeeded in his Project 56, setting a new world record of $56.88$ seconds in 2019.

For each event, Figure \ref{fig:results_swimtimes} shows these estimated ultimate times normalised by the corresponding current world records as of the beginning of 2019, ordered by increasing threshold swim-times.
For the vast majority of events, the ultimate time is $93$-$95$\% of the current world record. For the women's 50m butterfly and women's 1500m freestyle however, the current world record is very close to the ultimate time. In fact, approximately a $3$\% improvement would see these ultimate times being reached. This finding is not so surprising as these two world records correspond to the top two ranks, Sarah Sjostrom and Katie Ledecky from Figure \ref{fig:ranks_plot}. In comparison, the world record swim-time for the men's 100m free, which does not make the top 20 ranks, would require a $7$\% improvement to reach the ultimate time, suggesting this is the weakest of the current world records.   

\subsection{Expected new world record time}
\begin{figure}
\centering
\includegraphics[scale=1]{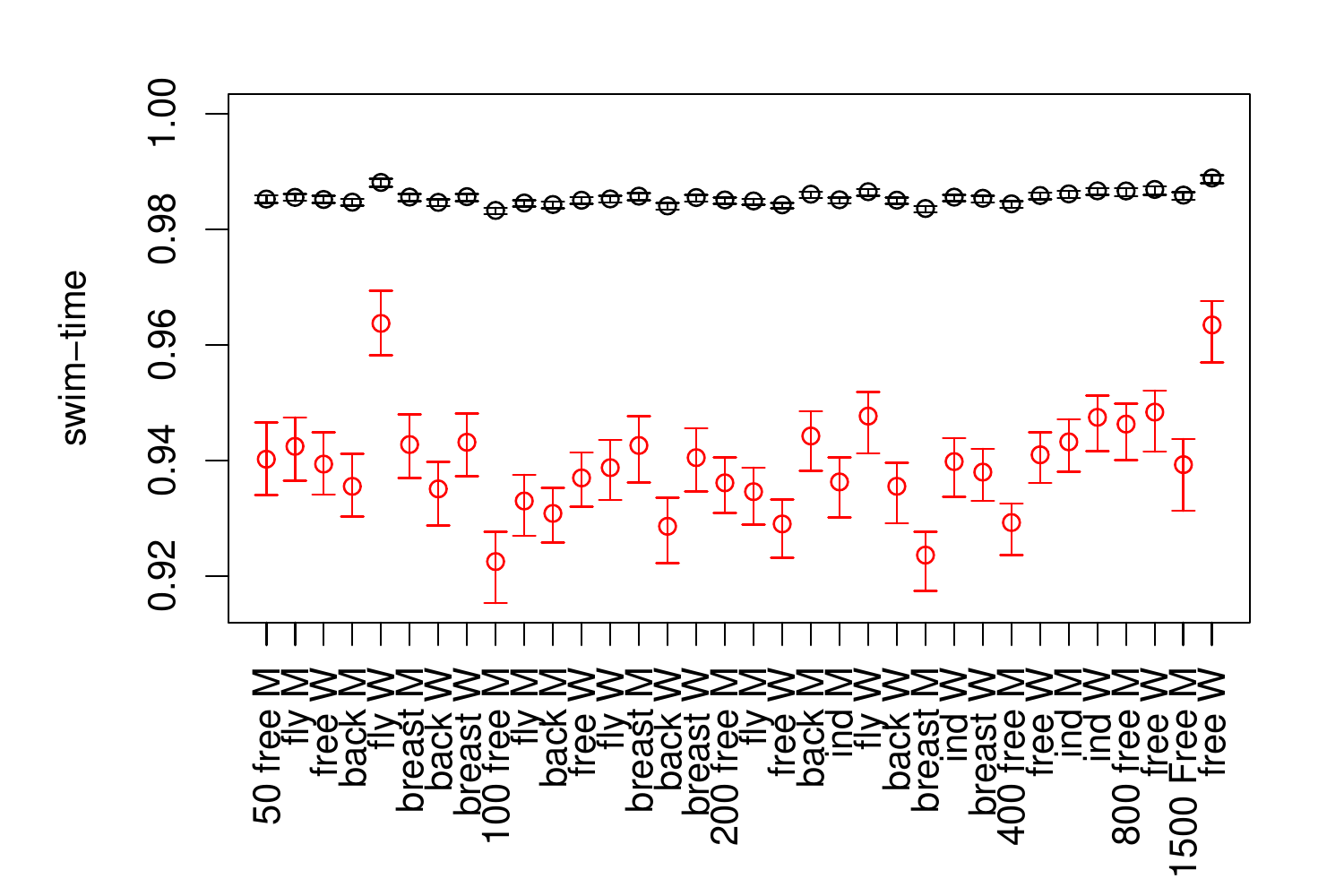}
\caption{The estimated expected next world record swim-time (upper black) and ultimate possible time (lower red) for each event the values are rescaled by world record as at the end of 2018, with 95 \% CI's from bootstrapped data sets.}
\label{fig:results_swimtimes}
\end{figure}

Let $X^{*(e)}_t$ be the random variable denoting the swim-time of a new world record in event $e$ at time $t$, then the distribution of $X^{*(e)}_t$ follows immediately from equations \eqref{eq:probExc} and \eqref{eq:GPd}, i.e., \begin{equation}
\label{eq:expected_rec}
\Pr\{X^{*(e)}_t > x \} = \Pr\{X^{(e)}_t > x | X^{(e)}_t > r_e \} = \bar{H}^{(e)}_{r_e}(x)
 =  \left[1+\xi\left(\frac{x-r_e}{\tilde{\sigma}^{(e)}_{r_e}}\right)\right]_+^{-\frac{1}{\xi}}, \textup{if}\; x>r_e,
\end{equation}
where $\tilde{\sigma}^{(e)}_{r_e} = \sigma^{(e)}_0 + \xi (r_e - \mu^{(e)}_0)$ and $r_e$ is the world record for event $e$ at the end of 2018 such that $r_e:= \max(\pmb{x}_e)$ where $\pmb{x}_e$ are all the observations in event $e$. Note that the right hand side of expression \eqref{eq:expected_rec} has no time dependency, since under model $\mathcal{M}_{7b}$ the distribution of times, conditional on being above threshold $u_e$, is time homogeneous for any given event $e$, see property \eqref{eq:stationarysigma}, therefore we drop the subscript $t$. From this the expected swim-time of the next world record in event $e$ is $$\mathbb{E}[X^{*(e)}] = \int_{r_e}^{x_{H,e}} x \frac{\diff H^{(e)}_{r_e}(x)}{\diff x} \diff x = r_e +\frac{\tilde{\sigma}_{r_e}^{(e)}}	{1-\xi}, \; \textup{if} \; \xi < 1.$$
Figure \ref{fig:results_swimtimes} shows the estimated expected swim-time of the next world record relative to the world record at the end of 2018, where events are ordered by increasing swim-time. Censoring is ignored in this calculation, as it would have such a negligible effect. The expected improvement varies only slightly between events, ranging from an expected improvement of $0.5$\% for Katie Ledecky's 1500m women's free performance, to a $0.9$\% for Cesar Cielo's 100m men's free performance. In events where the ultimate time is close to the current record, the expected next world record is also closer to the current record, and vice versa. 

The small variation between expected improvement is at first surprising, since it might be expected that `better' records, such as those of Katie Ledecky and Sarah Sjostrom, would be beaten by much smaller amounts. However, it is also likely that these records will take longer to be broken and so the improvements in training methods will be more significant by the time a new record is set, which may reduce the variation in the percentage improvement.

The confidence intervals here describe the confidence in the mean of the corresponding estimate, but it might also be interesting to determine the prediction interval, e.g., the $95$\% interval of possible swim-times that the next world record swim-time in event $e$ will be in. The predictive distribution of $\Pr\{X^{*(e)} < x \}$ can be found as follows: if $\{ \hat{\Theta}^{(i)}:\; i=1,\dots,n\}$ are the $n=240$ bootstrapped parameter estimates, where $\hat{\Theta}^{(i)}$ corresponds to the MLE's from simulated data set $i$, then for large $n$ and $x> r_e$ \begin{align*}
\Pr\{X^{*(e)} < x \} \approx &  \frac{1}{n}\sum_{i=1}^{n} \Pr\{X^{*(e)} < x | \hat{\Theta}^{(i)}\}\nonumber \\ 
=& \frac{1}{n} \sum_{i=1}^{n} H_{r_e}^{(e)}(x|\hat{\Theta}^{(i)})\nonumber\\
&=1- \frac{1}{n}\sum_{i=1}^{n}\left[1+\xi^{(i)}\left(\frac{x-r_e}{\tilde{\sigma}^{(i,e)}_{r_e}}\right)\right]_+^{-\frac{1}{\xi^{(i)}}},
\end{align*} where $\xi^{(i)}$ and $\tilde{\sigma}^{(i,e)}_{r_e}$ are the bootstrapped parameter estimates for $\xi$ and $\tilde{\sigma}^{(e)}_{r_e}$ corresponding to simulated data set $i$. Similar predictive distributions can be found for the other features of interest in Figures~\ref{fig:time_til} and \ref{fig:results_PMF}, as described in Sections \ref{sec:time_until} and \ref{sec:prob}.

\subsection{Time until world record is next set for an event}
\label{sec:time_until}
\begin{figure}
\centering
\includegraphics[scale=1]{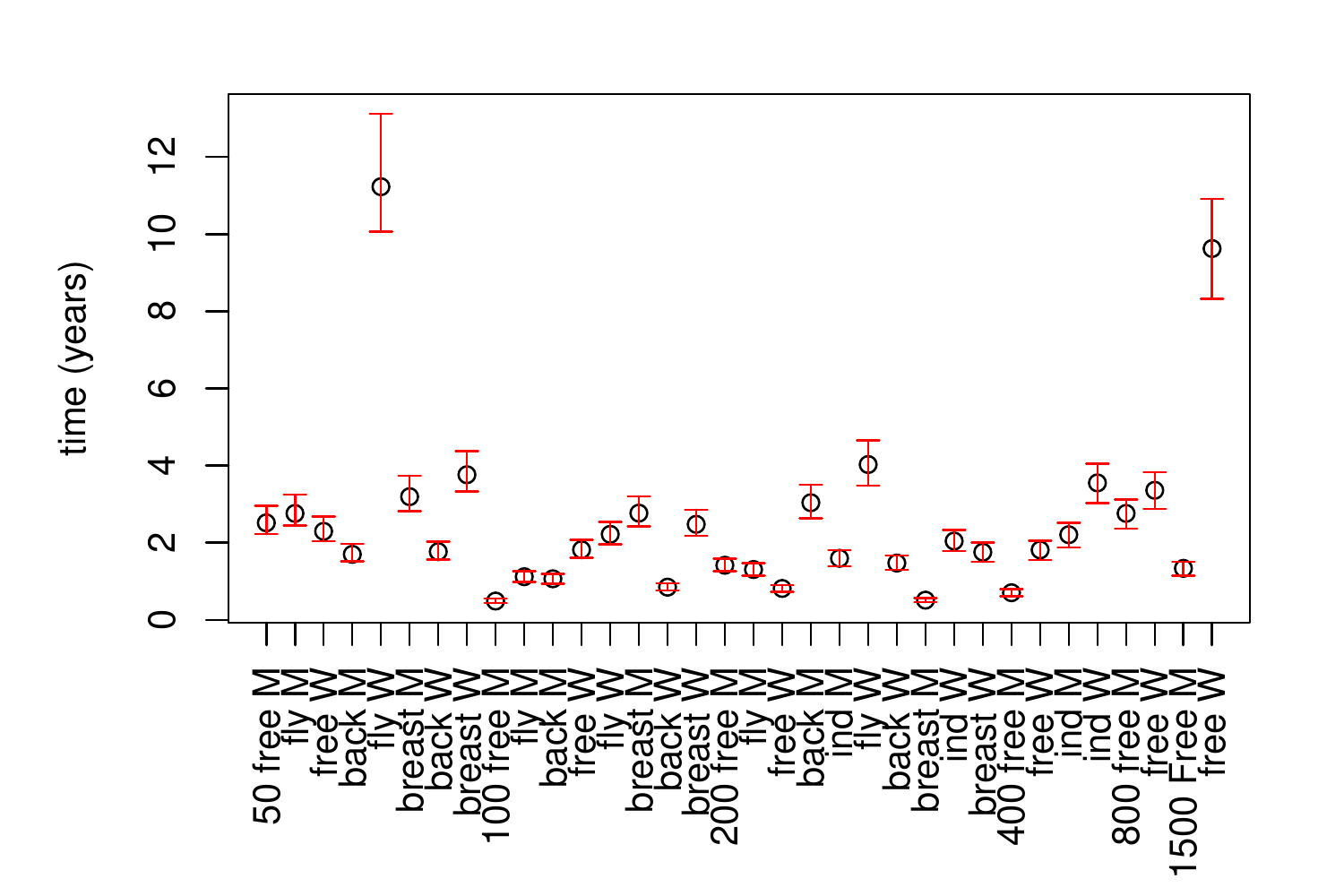}
\caption{The estimated expected time (in years) until the world record is broken with 95\% CI's from bootstrapped data sets.}
\label{fig:time_til}
\end{figure}

The distribution of time taken until a new world record is set in a particular event $e$ is of interest. Let $T^{(e)}$ be a random variable describing the time at which a new world record is next set in event $e \in E$. The probability $ F_{T^{(e)}}(t) = \Pr\{ T^{(e)} < t\}$ that a world record for event $e$ is set before some time $t$ can be found as follows. For current time 1, until a time $t$ $(t>1)$ there will be $N_t^{(e)}$ exceedances of the threshold $u_e$ in event $e$, and for the current record to be first broken after $t$ all of the $N_t^{(e)}$ observations need to be slower than the current record. Therefore, the following notation is introduced: let $X^{(e)}_{1:N_t^{(e)}} = \{X_i^{(e)}, i= 1,\dots, N_t^{(e)}\}$ where $X_i \overset{\text{iid}}{\sim} H_{u}^{(e)}$ and $H_{u}^{(e)}$ has GPd. Then $N_t^{(e)}$ has a Poisson distribution with mean $$\Lambda^{(e)}\left(\mathcal{A}_{(1,t),u}\right) = \int_{1}^{t} \left[1+\xi \left(\frac{u_e-\mu^{(e)}(y)}{\sigma^{(e)}(y)}\right)\right]^{-\frac{1}{\xi}}_+ \diff y,  $$ and the probability that a world record for event $e$ is set before $t$ is
\begin{align}
\label{eq:timedist}
F_{T^{(e)}}(t)&=1- \Pr\{T^{(e)}>t\} \nonumber\\
& = 1 - \sum_{m=0}^\infty \Pr\{\max(X^{(e)}_{1:N_t^{(e)}}) < r_e | N_t^{(e)}=m\} \Pr\{N_t^{(e)}=m\} \nonumber\\
&=1-\sum_{m=0}^\infty \left[H_{u}^{(e)}(r_e)\right]^m \left[\Lambda^{(e)}\left(\mathcal{A}_{(1,t),u}\right)\right]^m \exp\left[-\Lambda^{(e)}\left(\mathcal{A}_{(1,t),u}\right)\right]/m! \nonumber\\
&= 1- \exp\left[-\Lambda^{(e)}\left(\mathcal{A}_{(1,t),u}\right) \bar{H}^{(e)}_{u}(r_e) \right],
\end{align}
where the final equality follows from the power series expression for the exponential function. The density function for $T^{(e)}$, $f_{T^{(e)}}$, follows from equation \eqref{eq:timedist}, as
\begin{equation*}
f_{T^{(e)}}(t) = \left[1+\xi \left(\frac{u_e-\mu^{(e)}(t)}{\sigma^{(e)}(t)}\right)\right]^{-\frac{1}{\xi}}_+ \bar{H}_{u}^{(e)}(r_e) \exp\left[-\Lambda^{(e)}\left(\mathcal{A}_{(1,t),u}\right)\bar{H}_{u}^{(e)}(r_e)\right].
\end{equation*}
Then the expected time until a world record is next set in event $e$ is \begin{equation*}
\mathbb{E}\left[T^{(e)}\right] = \int_1^\infty t f_{T^{(e)}}(t) \diff t.
\end{equation*} Figure \ref{fig:time_til} shows these MLE's along with 95\% confidence intervals for $\mathbb{E}\left[T^{(e)}\right]$. It can be seen that almost all events are expected to have a new world record in the next 5 years. The longest estimated expected waiting times are again the times until Katie Ledecky's and Sarah Sjostrom's world records are broken, in the women's 1500m free and women's 50m fly respectively which correspond to the top two ranks of Figure \ref{fig:ranks_plot}, which both have expected waiting times of approximately 11 years. 

\subsection{Probability that a record is next set in a particular event}
\label{sec:prob}
\begin{figure}
\centering
\includegraphics[scale=1]{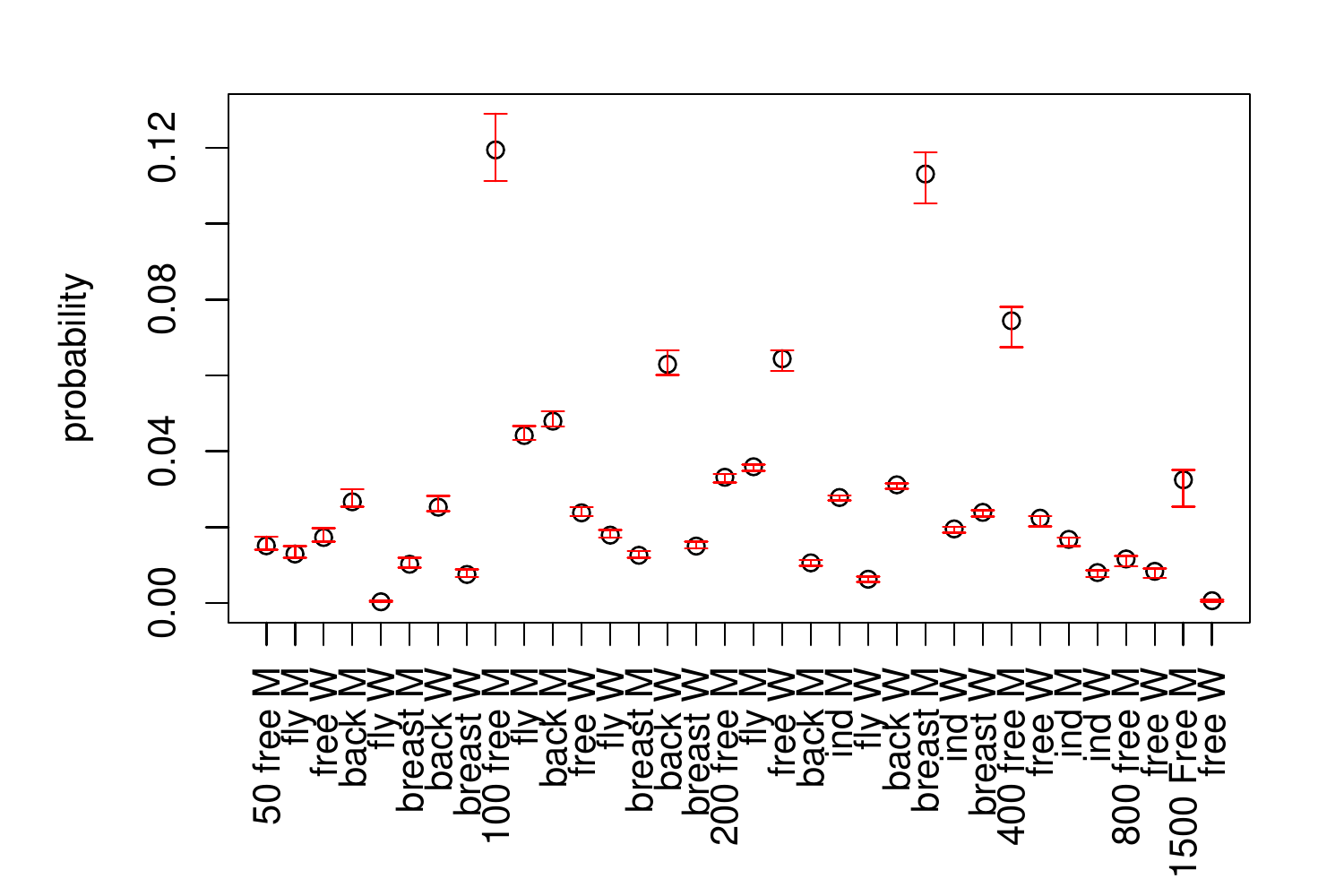}
\caption{Estimated probabilities that the next world record is set in a particular event, with $95$\% CI's from bootstrapped data sets.}
\label{fig:results_PMF}
\end{figure}

Now suppose that we wish the find the probability that the next event to have a world record that is broken is in event $e$. Let $T^{(-e)}$ be the random variable denoting the time taken for a world record to be set in any other event apart from $e$, i.e., $$ T^{(-e)} :=  \min_{k \in E \setminus \{e\} } \; \{T^{(k)}\}.$$ Then the probability that the next world record that is set is in event $e$ is given by
\begin{align*}
\Pr\{ & T^{(-e)} > T^{(e)} \} \nonumber\\
 &=\int_1^\infty \Pr\{T^{(-e)} > T^{(e)} | T^{(e)} = t\}\Pr\{T^{(e)}=t\} \diff t  \nonumber\\
&= \int_1^\infty  \prod_{k \in E \setminus \{e\}} \left\{ \exp\left[-\Lambda^{(k)}\left(\mathcal{A}_{(1,t),u}\right)\bar{H}^{(k)}_{u}(r_k)\right] \right\}  \nonumber\\
& \quad \left[1+\xi \left(\frac{u_e-\mu^{(e)}(t)}{\sigma^{(e)}(t)}\right)\right]^{-\frac{1}{\xi}}_+ \bar{H}_{u}^{(e)}(r_e)  \exp\left[-\Lambda^{(e)}\left(\mathcal{A}_{(1,t),u}\right)\bar{H}_{u}^{(e)}(r_e)\right] \diff t    \nonumber\\
&= \int_1^\infty  \left\{ \exp\left[-\sum_{k \in E}\Lambda^{(k)}\left(\mathcal{A}_{(1,t),u}\right)\bar{H}_{u}^{(k)}(r_e)\right] \right\} \left[1+\xi \left(\frac{u_e-\mu^{(e)}(t)}{\sigma^{(e)}(t)}\right)\right]^{-\frac{1}{\xi}}_+ \bar{H}_{u}^{(e)}(r_e) \diff t,
\end{align*}
where the second equality follows because
 \begin{equation}
\label{eq:prodtimedist}
\Pr\{T^{(-e)} > T^{(e)} | T^{(e)} = t\} = \prod_{k \in E \setminus \{e\}} \left\{ \exp\left[-(\Lambda^{(k)}\left(\mathcal{A}_{(1,t),u}\right)\bar{H}_{u}^{(k)}(r_k)\right] \right\}
\end{equation} due to the assumption of independence between swims in different events and the result derived in equation \eqref{eq:timedist} for a single event. Figure \ref{fig:results_PMF} shows these estimated probabilities with the previously identified `better' records having a lower probability of being broken next. The most likely record to be broken is the men's 100m free. 
The estimates of these probabilities using model $\mathcal{M}_{1b}$ was compared (not shown), and it has less variance between events.

\subsection{Adjusting Swim-Suit Influenced Times}
In 2010 Brazil's Cesar Cielo called for FINA to scrap any records set in the now-banned swim-suits, due to those records being much more difficult to break. Rather than this however, it is desirable to find a fair comparison between swim-times of those swimmers wearing a swim-suit and those not, and even construct a framework such that swim-times can be fairly compared with other future technological advancements.

Since the rank of a swim-time is based on the rate $R$ at which better observations occur, it is possible to adjust the swim-time for the use of a swim-suit. Let $x > u$ be a swim-time occurring at time $q$ during the swim-suit period i.e, $q \in S_{t_1} \cup S_{t_2}$, and $z$ is a swim-time occurring at the same time but as if it were not swam using a swim-suit. Then the swim-time correction from a recorded swim-time $x$ to an equivalent swim-time without the swim-suit $z$ is made by selecting $z$ such that the rate of exceeding $x$, $R$, and the corrected rate of exceeding $z$ without a swim-suit, $R_C$, are equal. That is, find $z$ as the solution to 
\begin{equation}
\label{eq:RCdefinition}
R\{X_{q}^{(e)} > x\} = R_C\{X_{q}^{(e)} > z\},
\end{equation}
where $R$ is defined in equation \eqref{eq:rankings_} and $R_C$ is defined by $$R_C\{X_{q}^{(e)} > z\} = \Pr\{X_{q}^{(e)} > z | X_{q}^{(e)} > u_e\} \Lambda^{(e)}_{C,q}\left(\mathcal{A}_{1,u}\right),$$
where $$ \Lambda^{(e)}_{C,q}\left(\mathcal{A}_{1,u}\right)=\left[1+\xi \left(\frac{u_e-\mu_C^{(e)}(q)}{\sigma_C^{(e)}(q)}\right)\right]_+^{\frac{1}{\xi}},$$ $\sigma_C^{(e)}(q) = \sigma_0^{(e)} + \xi \beta q$, and $ \mu_C^{(e)}(q) = \mu_0^{(e)} + \beta q$.
Thus, the adjusted swim-time $z$ is found via the solution to equation \eqref{eq:RCdefinition}, given as
\begin{equation*}
z = u_e + \frac{\tilde{\sigma}_u^{(e)}}{\xi} \left\{\frac{ \Lambda^{(e)}_{q}\left(\mathcal{A}_{1,u}\right)\bar{H}_u^{(e)}(x)} { \Lambda^{(e)}_{C,q}\left(\mathcal{A}_{1,u}\right)}-1 \right\}.
\end{equation*}
As an example, Cesar Cielo's 6th rank swim-time of 20.91s in the 50m freestyle in 2009 gets adjusted to 21.18 once the swim-suit effect is removed. The reverse can be found, that is the time a swimmer would have got, had they been wearing a swim-suit, e.g., Adam Peaty's current 100m breaststroke world record time of 56.88s gets adjusted to 56.25s with a swim-suit from 2008, and adjusted to 55.96 with a swim-suit from 2009, indicating that a ``Project 55'' could be achieved with just the addition of a swim-suit. By adjusting for technology in this way, it is possible to determine which current world records would still stand, had swim-suits never played a part. Table \ref{tab:adjusted_swim_times} shows those current world records set using swim-suits, and their estimated adjustments. Moreover, Table \ref{tab:adjusted_swim_times} shows what the world record would be, and who the world record holder would be, once the effect of swim-suits is removed. Out of the 10 world records which have been set by swimmers wearing swim-suits, only 2 would still stand today, Zige Liu's 200m fly world record, and Zhang Lin's 800m free world record. It is worth noting that the assumption that the most up-to-date technology available is always being used, is occasionally violated, for example, Phelps' 100m and 200m fly world records in 2009 were swam with the LZR Speedo suits from 2008. There can be additional complications when taking technology into account, such as Phelps' 400m  individual medley world record from 2008, in which  only the leg suit was worn. These issues could be addressed with the addition of explicit data about which technology was being used in a given swim.
\begin{table}[!htbp] \centering 
\begin{tabular}{@{\extracolsep{5pt}} cccccc} 
\\[-1.8ex]\hline 
\hline \\[-1.8ex] 
Event & WR swim & WR & AWR & NSWR & NSWR swim \\ 
\hline \\[-1.8ex] 
50 free M & Cielo (2009) & $20.91$ & $21.18$ & $\pmb{21.11}$ & \textbf{Proud (2018)} \\
100 free M & Cielo (2009) & $46.91$ & $47.99$ & $\pmb{47.04}$ & \textbf{McEvoy (2016)} \\  
100 fly M & Phelps (2009) & $49.82$ & $50.83$ & $\pmb{49.86}$ & \textbf{Dressel (2017)} \\ 
200 fly M & Phelps (2009) & $111.51$ & $113.33$ & $\pmb{112.71}$ & \textbf{Milak (2018)} \\ 
200 back M & Peirsol (2009) & $111.92$ & $113.47$ & $\pmb{112.96}$ & \textbf{Lochte (2011)} \\
200 free F & Pellegrini (2009) & $112.98$ & $114.99$ & $\pmb{113.61}$ & \textbf{Schmitt (2012)} \\
200 fly F & \textbf{Zige (2009)} & $121.81$ & $\pmb{123.38}$ & $124.06$ & Jiao (2012) \\ 
400 free M & Biedermann (2009) & $220.07$ & $223.13$ & $\pmb{220.08}$ & \textbf{Thorpe (2002)} \\ 
400 ind M & Phelps (2008) & $243.84$ & $245.72$ & $\pmb{245.18}$ & \textbf{Lochte (2012)} \\ 
800 free M & \textbf{Lin (2009)} & $452.12$ & $\pmb{455.31}$ & $458.57$ & Sun (2011) \\ 
\hline \\[-1.8ex] 
\end{tabular} 
\caption{World records (WR) set with swim-suits, the adjusted times (AWR), and the best corresponding non-swim-suit times (NSWR). ``Would-be'' world records and world record holders, after adjusting for swim-suits, are marked in bold.}
  \label{tab:adjusted_swim_times}
\end{table}

\section{Discussion}
\label{sec:Discussion}
Throughout this article, the swim-times are negated before being analysed so that we can use existing methodology for larger values. Alternatively, by analysing swim-\textit{speed}, \cite{gomes2019swimming} apply peaks-above-threshold methodology directly, since a smaller swim-time equates to a larger swim-speed. This raises the question of which transformation is best, and what classes of transformation give similar results. From limit \eqref{eq:probMax}, it can be seen that any linear transformations will be absorbed into the norming constants $a_n$ and $b_n$ so that inference is invariant for positive linear transformations. Conversely, \cite{wadsworth2010accounting} show that non-linear transformations lead to different results. \cite{wadsworth2010accounting} consider the class of Box-Cox transformations as part of the extreme value analysis with negating of the data and inversion to swim-speed as special cases. Thus a possible route for future research is to find the best Box-Cox parameter and to see if this changes in a systematic way over distance, gender and stroke.

Only the best time is recorded from each swimmer in a given event which, for cases where swimmers in the data set are still active, could lead to poor predictive performance. For example, let $X^{(w,e)}_t$ be the random variable denoting a swim-time by the current world record holder in event $e$ at time $t$, and $X^{(i,e)}_t$ be the random variable denoting a swim-time by another swimmer $i$ in event $e$ at time $t$, then the probability of a world record-holder setting a new personal best, and therefore new world record, is likely to be larger than the probability of any new swimmer setting a world record, such that $\Pr \{X^{(w,e)}_\tau > r_e \} > \Pr \{X^{(i,e)}_\tau > r_e \}$ for $\tau>t$. This could be accounted for by allowing more than one swim-time to be recorded per swimmer, however this gives rise to dependency between swim-times in the same event, and would need to be adjusted for.

Independence is assumed between swim-times for different strokes, genders and distances. This simplifying assumption may not be true when the same swimmer competes across many distances or strokes, meaning that the uncertainty of our estimates would be underestimated. Of the swim-times that exceed the thresholds $u_e$, i.e., for $e \in E$, the proportion of unique swimmers to total data points is around half, and so the effective sample size of independent swimmers in the data set will be less than the number of total data. In the case that there is perfect correlation between the same swimmer in separate events, then the effective sample size will be equal to the number of unique swimmers, approximately half the total data values, which means the variance could be underestimated by at most a factor of 2. This could be corrected for by estimating some inflation parameter $1 \leq \phi \leq 2$, such that the actual variance is equal to $\phi \textup{var}(\hat{\theta})$, where $\textup{var}(\hat{\theta})$ is the variance obtained by assuming complete independence between observations, see \cite{kent1982robust}. It may be necessary to use multivariate techniques in order to capture some of the correlation between data points resulting from the same swimmer in different competitions \citep{doi:10.1080/15598608.2012.695702}. 

There are extra sources of uncertainty not accounted for. Quantifying the uncertainty due to the choice of threshold is not considered, since a single threshold selection approach is used, as is common in the extreme value theory literature \citep{scarrott2012review}. However this uncertainty could be quantified by using the cross-validatory technique of \cite{northrop2017cross}. Also, since the analysis is performed in a frequentist framework, only parameter uncertainty is considered, however when predicting future events such as the time until a new world record is set in a particular event, it is also valuable to consider the predictive uncertainty.
This could be accounted for by moving to a Bayesian framework, and carrying out parameter estimates via Markov chain Monte Carlo with a prior on the spline roughness penalty.

The constant evolution of the para-swimming classification system is testament to the challenge of creating fair competition in disability swimming. The number of classifications itself is open to debate, with too many classifications resulting in too few swimmers in each classification and therefore a drop in competitiveness, and too few classifications resulting in bias such that there is unfair differences between swimmer's physical limitations within the same class. Of course, this problem stems from the discrete nature of the classification system, but a model of the type presented in this article would allow for a continuous ``classification variable'' which pools across disability, to allow fair competition over all disability types and comparison between disabilities.
In a similar way, this model could allow for more fair comparison with transgender swimmers. Regulations around transgender athletes in sports is a controversial topic, with the regulations being changed again for the upcoming 2020 Olympic Games. This controversy largely arises due to determining whether a transgender athlete should compete in the men's or women's event, and is determined on a case by case basis. However, our type of covariate model can allow for a more fluid description of gender, since the adjustment or categorisation is determined simply by the threshold time $u_e$ which can easily be modelled as continuous across events or gender status. In addition, cases of unusual testosterone levels can be dealt with in the same way. 
In junior swimming, because of the discretisation of age groups, some swimmers can be almost a whole year younger than others in the same competition, which creates an unfair disadvantage. The same idea of a continuous scale for age groups would allow for fair comparison of `age-adjusted' swim-times.
Ultimately, it is possible to have a global model which fairly compares swimmers of all genders and disabilities, and even junior swimmers, across different events.


\linespread{1}
\footnotesize{

\subsection*{Acknowledgements}
Spearing gratefully acknowledges funding of the EPSRC funded STOR-i centre for doctoral training
(grant number EP/L015692/1), and ATASS sports. We thank the referees for their helpful comments.
\linespread{1}
\bibliography{Swimming}	\bibliographystyle{apalike}

\begin{thebibliography}{}

\bibitem[Adam and Tawn, 2012]{doi:10.1080/15598608.2012.695702}
Adam, M.~B. and Tawn, J.~A. (2012).
\newblock Bivariate extreme analysis of {O}lympic swimming data.
\newblock {\em Journal of Statistical Theory and Practice}, 6(3):510--523.

\bibitem[Blest, 1996]{blest1996lower}
Blest, D.~C. (1996).
\newblock Lower bounds for athletic performance.
\newblock {\em Journal of the Royal Statistical Society: Series D (The
  Statistician)}, 45(2):243--253.

\bibitem[Coles, 2001]{coles2001introduction}
Coles (2001).
\newblock {\em An Introduction to Statistical Modeling of Extreme Values},
  volume 208.
\newblock Springer London.

\bibitem[Davison and Smith, 1990]{davison1990models}
Davison, A.~C. and Smith, R.~L. (1990).
\newblock Models for exceedances over high thresholds.
\newblock {\em Journal of the Royal Statistical Society: Series B
  (Methodological)}, 52(3):393--425.

\bibitem[De~Boor, 1978]{de1978practical}
De~Boor, C. (1978).
\newblock {\em A Practical Guide to Splines}, volume~27.
\newblock Springer-Verlag New York.

\bibitem[Eilers and Marx, 1996]{eilers1996flexible}
Eilers, P.~H. and Marx, B.~D. (1996).
\newblock Flexible smoothing with {B}-splines and penalties.
\newblock {\em Statistical Science}, 1:89--102.

\bibitem[Ewans and Jonathan, 2008]{ewans2008effect}
Ewans, K. and Jonathan, P. (2008).
\newblock The effect of directionality on northern {N}orth {S}ea extreme wave
  design criteria.
\newblock {\em Journal of Offshore Mechanics and Arctic Engineering},
  130(4):041604.

\bibitem[Foster et~al., 2012]{foster2012influence}
Foster, L., James, D., and Haake, S. (2012).
\newblock Influence of full body swimsuits on competitive performance.
\newblock {\em Procedia engineering}, 34:712--717.

\bibitem[Gomes and Henriques-Rodrigues, 2019]{gomes2019swimming}
Gomes, D.~T. and Henriques-Rodrigues, L. (2019).
\newblock Swimming performance index based on extreme value theory.
\newblock {\em International Journal of Sports Science \& Coaching},
  14(1):51--62.

\bibitem[Huub and Trultens, 2005]{toussaint122005biomechanical}
Huub, T. and Trultens, M. (2005).
\newblock Biomechanical aspects of peak performance in human swimming.
\newblock {\em Animal Biology}, 55(1):17--40.

\bibitem[Kent, 1982]{kent1982robust}
Kent, J.~T. (1982).
\newblock Robust properties of likelihood ratio tests.
\newblock {\em Biometrika}, 69(1):19--27.

\bibitem[Moria et~al., 2011]{moria2011aero}
Moria, H., Chowdhury, H., Alam, F., and Subic, A. (2011).
\newblock Aero/hydrodynamic study of {S}peedo {LZR}, {TYR} {S}ayonara and
  {B}lueseventy pointzero 3 swimsuits.
\newblock {\em Jordan Journal of Mechanical and Industrial Engineering},
  5(1):83--88.

\bibitem[Nevill et~al., 2007]{nevill2007there}
Nevill, A.~M., Whyte, G.~P., Holder, R.~L., and Peyrebrune, M. (2007).
\newblock Are there limits to swimming world records?
\newblock {\em International Journal of Sports Medicine}, 28(12):1012--1017.

\bibitem[Northrop et~al., 2017]{northrop2017cross}
Northrop, P.~J., Attalides, N., and Jonathan, P. (2017).
\newblock Cross-validatory extreme value threshold selection and uncertainty
  with application to ocean storm severity.
\newblock {\em Journal of the Royal Statistical Society: Series C (Applied
  Statistics)}, 66(1):93--120.

\bibitem[Ramsay, 1988]{ramsay1988monotone}
Ramsay, J.~O. (1988).
\newblock Monotone regression splines in action.
\newblock {\em Statistical Science}, 3(4):425--441.

\bibitem[Riegel, 1981]{riegel1981athletic}
Riegel, P.~S. (1981).
\newblock Athletic records and human endurance: A time-vs.-distance equation
  describing world-record performances may be used to compare the relative
  endurance capabilities of various groups of people.
\newblock {\em American Scientist}, 69(3):285--290.

\bibitem[Robinson and Tawn, 1995]{robinson1995statistics}
Robinson, M.~E. and Tawn, J.~A. (1995).
\newblock Statistics for exceptional athletics records.
\newblock {\em Journal of the Royal Statistical Society: Series C (Applied
  Statistics)}, 44(4):499--511.

\bibitem[Scarrott and MacDonald, 2012]{scarrott2012review}
Scarrott, C. and MacDonald, A. (2012).
\newblock A review of extreme value threshold estimation and uncertainty
  quantification.
\newblock {\em REVSTAT--Statistical Journal}, 10(1):33--60.

\bibitem[Shibata, 1989]{shibata1989statistical}
Shibata, R. (1989).
\newblock Statistical aspects of model selection.
\newblock In {\em From {D}ata to {M}odel}, pages 215--240. Springer, Berlin,
  Heidelberg.

\bibitem[Shipley, 2009]{shipley2009fina}
Shipley, A. (2009).
\newblock F{INA} opts to ban all high-tech swimsuits.
\newblock {\em Reach for the Wall. com}, 24.

\bibitem[Smith, 1989]{smith1989extreme}
Smith, R.~L. (1989).
\newblock Extreme value analysis of environmental time series: an application
  to trend detection in ground-level ozone.
\newblock {\em Statistical Science}, 4(4):367--377.

\bibitem[Stephenson and Tawn, 2013]{stephenson2013determining}
Stephenson, A.~G. and Tawn, J.~A. (2013).
\newblock Determining the best track performances of all time using a
  conceptual population model for athletics records.
\newblock {\em Journal of Quantitative Analysis in Sports}, 9(1):67--76.

\bibitem[Strand and Boes, 1998]{strand1998modeling}
Strand, M. and Boes, D. (1998).
\newblock Modeling road racing times of competitive recreational runners using
  extreme value theory.
\newblock {\em The American Statistician}, 52(3):205--210.

\bibitem[Sylvan~Katz and Katz, 1999]{sylvan1999power}
Sylvan~Katz, J. and Katz, L. (1999).
\newblock Power laws and athletic performance.
\newblock {\em Journal of Sports Sciences}, 17(6):467--476.

\bibitem[Tawn, 1988]{tawn1988bivariate}
Tawn, J.~A. (1988).
\newblock Bivariate extreme value theory: models and estimation.
\newblock {\em Biometrika}, 75(3):397--415.

\bibitem[Wadsworth et~al., 2010]{wadsworth2010accounting}
Wadsworth, J., Tawn, J., and Jonathan, P.~e. (2010).
\newblock Accounting for choice of measurement scale in extreme value modeling.
\newblock {\em The Annals of Applied Statistics}, 4(3):1558--1578.

\end{thebibliography}
\begin{appendices}
\section{Spline Construction}\label{sec:appendSpline}
Let $B_k^d(x)$ be the value of the $k^{th}$ $d$ degree B-spline basis function at a point $x$, where $k = \{1,\dots,q\}$, $q \in \mathbb{Z}^+$, and $x_k$ denotes the $k^{th}$ knot, such that $B_k^d(x)$ is strictly positive within the region $x_{k} < x < x_{k+d}$. The exact form of the splines can be formed recursively from 0 degree basis splines. Note that 0 degree splines are trivial to form, described as step-functions over the region of each knot such that $$B_k^0(x) = \left\{\begin{matrix}
1,& \; x_k \leq x < x_{k+1}, \\
0, &\; \textup{otherwise.} 
\end{matrix} \right.$$
Then, using the formula \citep{de1978practical} for $d\geq 1$ $$B_k^{d+1}(x) = \frac{x - x_{k-(d+1)}}{x_{k-1}-x_{k-(d+1)}}B_k^d(x) + \frac{x_k - x}{x_{k}-x_{k-d}}B_{k+1}^d(x),$$ higher degree B-splines are formed. Figure \ref{fig: Bsplines} shows splines of degrees $d=1,2,3$. It can be seen that as the degree of the basis function increases, the function becomes smoother and has a larger range.
\begin{figure}
\centering
\includegraphics[scale=.7]{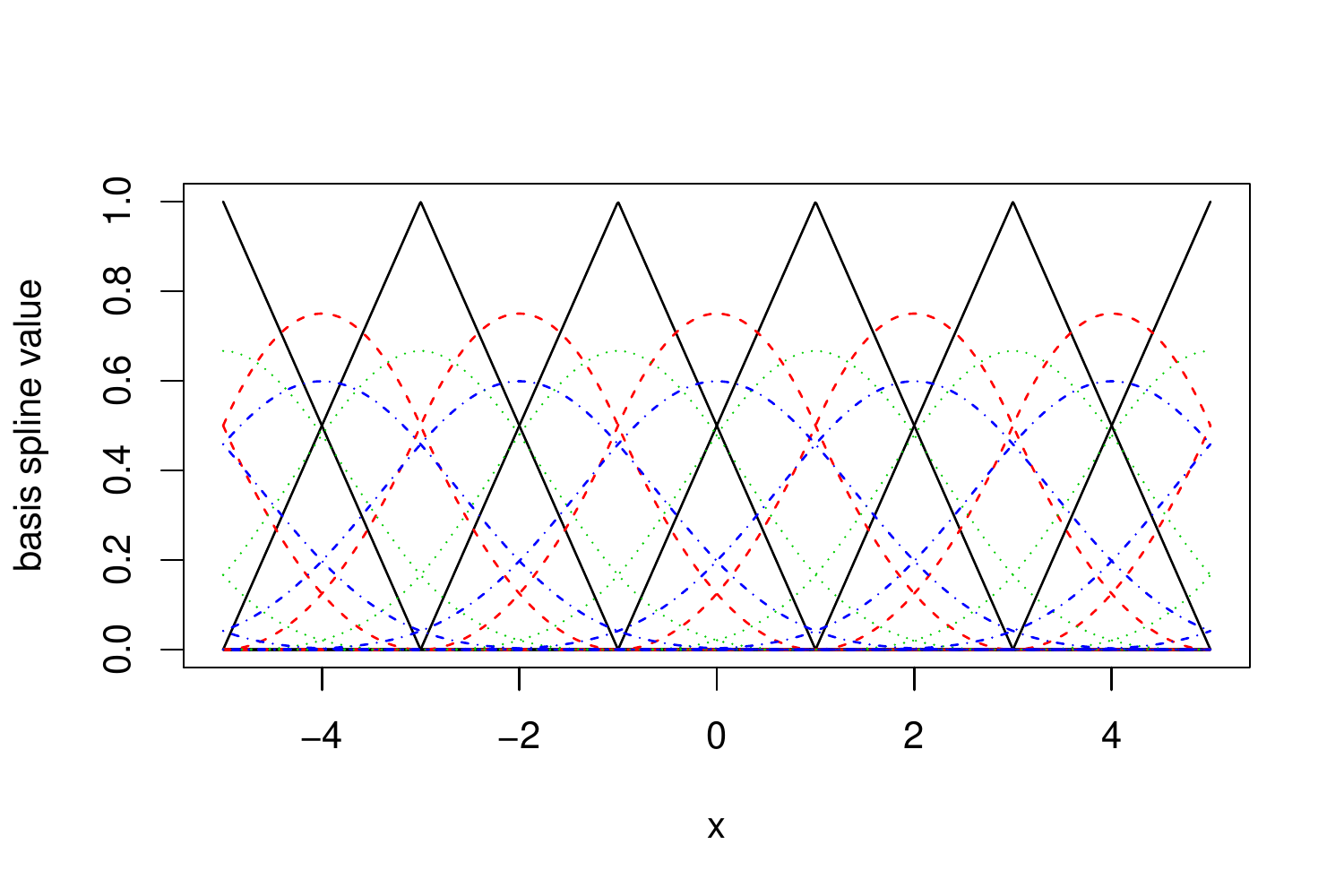}
\caption{Basis spline functions $B_k^d(x)$ with degree $d$: degree 1 (black solid), 2 (red dashed), 3 (green dotted), and 4 (blue dot-dashed), and knots are spaced at integer values.}
\label{fig: Bsplines}
\end{figure}
The spline function $Y(x)$ is then constructed as 
$$Y(x) = \sum_{k=1}^q a_k B^{d}_k(x)$$ where $a_k$ is the $k^{\textup{th}}$ B-spline coefficient, and $\pmb{a} = \{a_i: i=1,\dots,q\}$ is the coefficient vector. Generally, $q$ is chosen to be large, such that the fitted curve shows more variation than can be justified by the data. To reduce this variation, a penalty on the finite differences of adjacent coefficients of \cite{eilers1996flexible} is used. The penalty is governed by $\phi \pmb{a}' P \pmb{a}$, where $P \in \mathbb{R}^{q\times q}$ is the penalty matrix, and $\phi > 0$ determines the amount of penalisation. The choice of $P$ is based on some prior belief of the shape of the data. The penalty matrix used was a second order, such that

\begin{equation*}
P = \begin{bmatrix}
  1 & -2 & 1 & 0 & 0 & \dots & 0 & 0 & 0 \\ 
  -2 & 5 & -4 & 1 & 0 & \dots & 0 & 0 & 0 \\ 
  1 & -4 & 6 & -4 & 1 & \dots & 0 & 0 & 0 \\ 
  0 & 1 & -4 & 6 & -4 &\dots & 0 & 0 & 0 \\ 
  
  \vdots & \vdots & \vdots & \vdots & \vdots & \ddots & \vdots & \vdots & \vdots \\ 
  
  0 & 0 & 0 & 1 & -4 & \dots & -4 & 1 & 0 \\ 
  0 & 0 & 0 & 0 & 1 & \dots & 6 & -4 & 1 \\ 
  0 & 0 & 0 & 0 & 0 & \dots & -4 & 5 & -2 \\ 
  0 & 0 & 0 & 0 & 0 & \dots & 1 & -2 & 1 \\ 
  \end{bmatrix},
\end{equation*}
which penalises a large second derivative, thus penalising fits that depart from linearity.
\end{appendices}

\end{document}